\documentclass[12pt]{iopart}
\usepackage{amssymb}
\usepackage{epsf}
\usepackage{amsgen}
\usepackage{amsfonts}
\usepackage{amsbsy}
\newcommand{\krig}[1]{\stackrel{\circ}{#1}}
\begin{document}

\vspace{-2cm}

\hfill {\tiny{FZJ-IKP(TH)-01-12}}

\vspace{2cm}

\jl{4}

\topical{Axial structure of the nucleon}

\author{V\'eronique Bernard\dag, Latifa Elouadrhiri\ddag, Ulf-G Mei{\ss}ner\S}

\address{\dag\ Universit\'e Louis Pasteur de Strasbourg, Groupe
de Physique Th\'eorique, F-67084 Strasbourg Cedex 2, France}

\address{\ddag\ Thomas Jefferson National Accelerator Facility,
Newport News, VA 23606, USA}

\address{\S\ Forschungszentrum J\"ulich, Institut f\"ur Kernphysik
  (Theorie), D-52425 J\"ulich, Germany}

\hspace{1.7cm}{\small E-mail: bernard@lpt6.u-strasbg.fr, latifa@jlab.org,
u.meissner@fz-juelich.de}

\begin{abstract}
We review the current status of experimental and theoretical
understanding of the axial nucleon structure at low and moderate
energies. Topics considered include (quasi)elastic (anti)neutrino--nucleon
scattering, charged pion electroproduction off nucleons and ordinary
as well as radiative muon capture on the proton.
\end{abstract}


\submitted


\section{Introduction}
\label{sec:intro}
                    
The structure of the nucleon in the non-perturbative regime of QCD can be
characterized by a collection of scales, related to the probe  which is used
to study this basic building block of atomic nuclei. The typical nucleon
size as deduced from electron scattering experiments is about 0.85~fm, whereas
model--dependent analyses of proton--antiproton annihilation lead to a baryon
charge radius of about 0.5~fm.  An intermediate scale is set by the weak
charged currents, which allows to measure the axial charge radius,
$r_A \simeq 0.65$~fm,  and related moments. For a more detailed discussion
of these various scales and how they can be understood in a simple model,
see reference~\cite{UGM}. Here, we will be concerned with nucleon matrix
elements of the axial current, giving a status report on our current
knowledge about the pertinent coupling constants and form factors, with
particular emphasis on the theoretical methods which are used to extract
this information from experiment. Our basic object is the QCD axial current, 
expressed in terms of the light quark fields
\begin{equation}
A_\mu^a = \bar{q} \, \gamma_\mu \, \gamma_5 \, T^a \, q~,
\end{equation}
where for the two and three flavor case, the pertinent quark fields,
generators and flavor indices are given by
\begin{equation}
\begin{tabular}{r|ll}
\ms
 & SU(2) \qquad & SU(3) \qquad \nonumber \\
\hline
$q$ \quad & $\left(\begin{array}{c} u \\ d \end{array} \right)$  
& $\left(\begin{array}{c} u \\ d \\ s \end{array} \right)$
\nonumber \\
$T^a$ \quad& $\tau^a$ / 2  & $\lambda^a$ / 2 \nonumber \\
$a$  \quad & $1,2,3$ & $1, \ldots,8$ \\
\end{tabular}
\end{equation}
with $\tau^a$ the conventional Pauli isospin matrices and the $\lambda^a$ 
are Gell-Mann's SU(3) matrices. Note that we do not consider the singlet
components corresponding to $a=0$ in what follows. To be specific,
consider the matrix--element of the SU(2) isovector axial quark
current between nucleon states, 
\begin{equation}
\label{current}
\langle N(p') \, | A_\mu^a \, | N(p)\rangle = \bar{u}(p') \,
\left[ \gamma_\mu \, G_A (t) + {(p' -p)_\mu \over 2m}\,
G_P (t) \, \right] \, \gamma_5 {\tau^a \over 2} \, u(p)~,
\end{equation}
with $t=(p'-p)^2$ the invariant momentum transfer squared and
$m = (m_p + m_n)/2$ the nucleon mass. Note that in this review, we mostly consider
the isospin symmetric case and only differentiate between the proton ($m_p)$
and the neutron ($m_n)$ masses in kinematical factors.  The form
of equation~(\ref{current}) follows from  Lorentz invariance, isospin 
conservation and the discrete symmetries C, P and T and the absence
of second class currents~\cite{wein2}, which is consistent with experimental
information, see e.g. reference~\cite{MMK}.  $G_A (t)$ is called the nucleon
axial form factor and $G_P (t)$ the induced pseudoscalar form factor.
Consider first the axial form factor, which probes the spin--isospin
distribution of the nucleon (since in a non--relativistic language, this
is nothing but the matrix--element of the Gamov--Teller operator 
$\boldsymbol{\sigma} \, \boldsymbol{\tau}$). 
The axial form factor admits the following expansion for small 
momentum transfer
\begin{equation}
\label{taylor}
G_A (t) = g_A \, \left( 1 + \frac{1}{6} \langle r_A^2 \rangle \, t +
\Or(t^2) \right)~,
\end{equation}
with $g_A$ the axial--vector coupling constant, $g_A = 1.2673 \pm 
0.0035$~\cite{PDG},
and $\langle r_A^2 \rangle^{1/2}$ the nucleon axial radius. The
experimental determinations  and theoretical understanding of this
fundamental low--energy observable will constitute a main part of
this review. The axial--vector coupling constant is measured in
(polarized) neutron beta--decay, for recent references see e.g.
\cite{Reich,Abele}. Here, we will not discuss the
determinations of $g_A$ and of the related CKM matrix--element
$V_{ud}$, for a  review see \cite{TH}. 
 For low and moderate momentum transfer, say $|t| \leq
1\,$GeV$^2$, the axial form factor can be represented by a dipole
fit
\begin{equation}
\label{dipole}
G_A (t) =  \frac{g_A}{\left( 1 -  t/ M_A^2 \right)^{2}}~,
\end{equation}
in terms of one adjustable parameter, $M_A$, the so--called axial mass (or
sometimes dipole mass). Therefore, one can express the axial 
radius in terms of the dipole mass
\begin{equation}
\label{radmass}
\langle r_A^2 \rangle 
= {6\over g_A}\,
{{d}G_{\mathrm{A}}(t)\over{d}t}\,
\biggl\vert_{t=0}
= \frac{12}{M_A^2}~.
\end{equation}
The induced pseudoscalar form factor is believed to be understood in terms
of pion pole dominance and we will discuss the theoretical status concerning
the corrections to this leading order term. Most measurement of this form
factor stem from ordinary muon capture on the proton, $\mu^- + p \to \nu_\mu + n$.
More precisely, in that reaction one measures the induced pseudoscalar coupling
constant $g_P$,
\begin{equation}
g_p = \frac{M_\mu}{2m} \, G_P(t=-0.88M^2_\mu)~,
\end{equation}
with $M_\mu$ the muon mass.  This coupling constant is nothing but the value
of the induced pseudoscalar form factor $G_P (t)$ at the four--momentum
transfer for muon capture by the proton at rest,
\begin{equation}
t = - M_\mu^2 \, \left[ 1 - {(M_\mu+m_p)^2 - m_n^2 \over M_\mu (M_\mu + m_p)}
\right] = -0.88 \, M^2_\mu~. 
\end{equation}
In what follows, we will review these determinations and also discuss 
radiative muon capture as well as pion electroproduction which allow
to investigate the momentum dependence of $G_P (t)$.

\medskip\noindent
The material is organized in the following way. In section~\ref{sec:data},
we collect the available data on the axial form factor $G_A (t)$, the induced
pseudoscalar coupling constant $g_P$ and the corresponding form factor $G_P (t)$.
Section~\ref{sec:ax} is devoted to the theoretical methods which allow to 
determine the axial form factor and the axial radius of the nucleon as well
as the pion charge radius from (anti)neutrino scattering and pion electroproduction 
data. The QCD analysis of the induced pseudoscalar form factor and the methods
to extract it from ordinary and radiative muon capture are discussed in 
section~\ref{sec:gp}. Section~\ref{sec:GTR} is devoted to a discussion of 
certain axial matrix elements in the baryon octet pertinent to the quark
mass analysis. A short summary and outlook are given in section~\ref{sec:summ}.
Before proceeding, we point out that in this review {\em no} model calculations
of the matrix elements are reviewed, as illustrative as they might be. Our emphasis
is on applying a model--independent  approach tightly constrained by the symmetries
of the Standard Model to determine the axial structure of the nucleon in the
non-perturbative regime.

\section{Data overview}
\label{sec:data}

In this section, we will give a brief review of published data on the
axial and induced pseudoscalar form factors. The methods which lead
to these results will be presented and critically discussed in the
following sections. At this point, we only wish to point out that
systematical errors for certain processes have certainly been underestimated,
but here we simply collect the data as they are available.

\subsection{Data on the axial form factor}
\label{sec:axdata}

\begin{figure}[htb]
   \epsfysize=7cm
   \epsffile{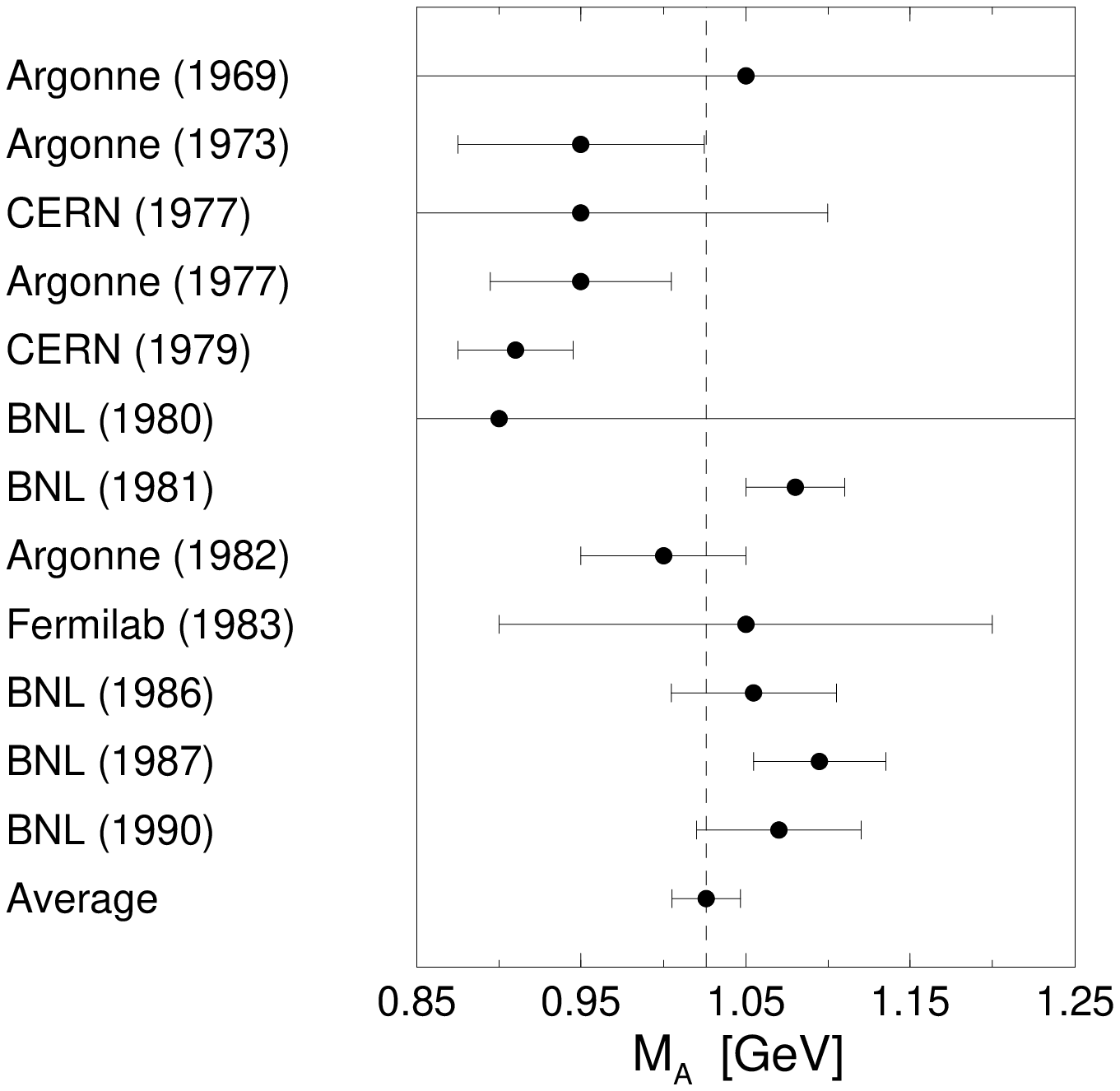}
\vspace{-7.0cm}
\hfill
   \epsfysize=7cm
   \epsffile{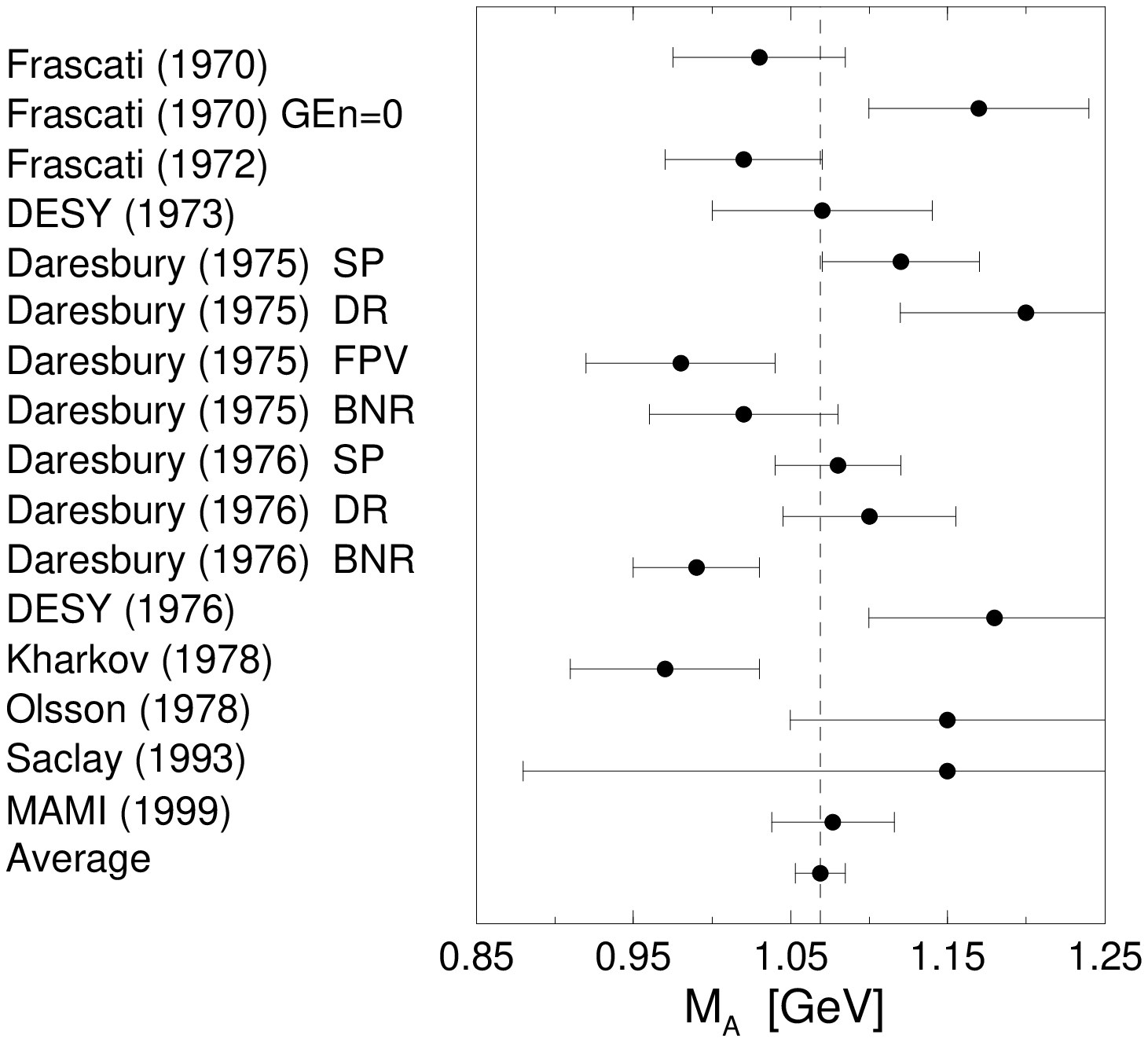}
   \vspace{.2cm} 
 \caption{\label{fig:axmass}
  Axial mass $M_A$ extractions. Left panel: From
 (quasi)elastic neutrino and antineutrino scattering experiments.
 The weighted average is $M_A=(1.026\pm 0.021)\,\mathrm{GeV}$.
 Right panel: From charged pion electroproduction experiments.
The weighted average 
is $M_A=(1.069\pm 0.016)\,\mathrm{GeV}$.
Note that value for the MAMI experiment contains both the statistical and
systematical uncertainty; for other values the systematical
errors were not explicitly given. The labels SP, DR, FPV and BNR
refer to different methods evaluating the corrections beyond the
soft pion limit as explained in the text.
}
\end{figure}

There are  two methods to determine the  axial
form factor of the nucleon, namely (anti)neutrino scattering
off protons or nuclei and charged pion electroproduction. 
We consider first the experimental data coming
from measurements of (quasi)elastic (anti)neutrino scattering
off protons \cite{fanourakis,ahrens1,ahrens2}, off deuterons 
\cite{barish}-\cite{kitagaki2}
and other nuclei (Al, Fe) \cite{holder,kustom}
or composite targets like freon 
\cite{perkins}-\cite{armenise}
and propane \cite{armenise,budagov}.  
In the left panel of figure~\ref{fig:axmass} we show the available values 
for the axial mass $M_A$ obtained from neutrino scattering
experiments. 
\noindent
As pointed out in \cite{MAMI}, 
references~\cite{holder,perkins,orkin,budagov} reported severe
uncertainties in either knowledge of the incident neutrino flux or
reliability of the theoretical input needed to subtract the background
from genuine elastic events (both of which gradually improved in
subsequent experiments).  The values derived in these papers fall well 
outside the most probable range of values known today and exhibit very large
statistical and systematical errors.  Following the data selection
criteria of the Particle Data Group \cite{PDG}, they were excluded 
from this compilation. In all cases, the axial form factor data were 
parameterized in terms of a dipole, the resulting world average is
\begin{equation}
\label{axmnu}
M_A=(1.026\pm 0.021)\,\mathrm{GeV} \quad \mathrm{(neutrino~~scattering)}~.
\end{equation}
The other determinations of the axial form factor are based
on the analysis of  charged pion electroproduction
off protons, see references~\cite{MAMI}\cite{amaldi1}-\cite{choiphd},
slightly above the pion production threshold (note that the MAMI
measurement is presently extended~\cite{MAMIax} to lower momentum transfer and to
check the cross section at the highest $Q^2$ point reported in \cite{MAMI}).
Such type of analysis is more involved. It starts from the low--energy theorem
of Nambu, Luri\'e and Shrauner \cite{NLS1,NLS2}
for the electric dipole amplitude $E_{0+}^{(-)}$ at threshold,
valid for soft pions, i.e. pions with vanishing four--momentum.
Model--dependent corrections (so--called hard pion corrections) were
developed to connect the low-energy theorem to the data, that is to the
real world with a finite pion mass, see 
references~\cite{nambu}-\cite{BNR2},
labeled SP, FPV, DR and BNR, respectively.
For a given model, the values of the axial mass were determined
from the slopes of the angle--integrated differential
electroproduction cross sections at threshold.
The results of various measurements and theoretical approaches
are shown in the right panel of  figure~\ref{fig:axmass}. 
Note again that references \cite{bloom,nambu} were omitted from 
the fit for lack of reasonable compatibility with the other results.
\begin{figure}[htb]
   \epsfysize=8cm
   \centerline{\epsffile{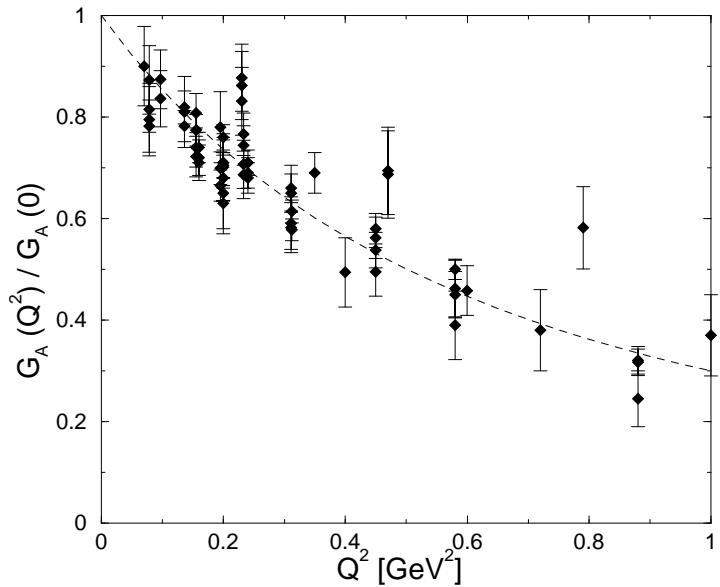}}
   \vspace{.2cm} 
  \parbox{15cm}{\caption{\label{alldata}
  Experimental data for the normalized axial form factor extracted
  from pion electroproduction experiments in the threshold region.
  Note that for the experiments where various theoretical models were
  used in the analysis to extract $G_A$, all results are shown.
  For orientation, the dashed line shows a  dipole fit  with an
  axial mass $M_A = 1.1\,$GeV.
  }}
\end{figure}
\noindent
In figure~\ref{alldata} we have collected the various electroproduction
data, in comparison to a dipole fit with $M_A =1.1\,$GeV. Again, at
various values of the momentum transfer one sees two or three data
points, these show the model--dependence due to the applied hard
pions corrections. The resulting world average of the dipole masses
collected in  figure~\ref{fig:axmass} is
\begin{equation}
\label{axmel}
M_A=(1.069\pm 0.016)\,\mathrm{GeV} \quad \mathrm{(electroproduction)}~.
\end{equation}
Although some of these results have large uncertainties, the weighted
average for the axial mass determinations from neutrino scattering
and pion electroproduction are quite precise. In particular, we notice 
an {\em axial mass discrepancy}, i.e. the so determined axial masses
differ significantly,
\begin{equation}
\label{axdis}
\Delta M_A=(0.043\pm 0.026)\,\mathrm{GeV}~,
\end{equation}
which translates into an axial radius difference of about 5\%, i.e.
the axial radius of the nucleon determined from neutrino scattering data
appears bigger than the one from electroproduction, 
$\langle r_A^2 \rangle_{\nu-{\rm scatt.}}^{1/2} = (0.666 \pm 0.014)\,$fm
versus $\langle r_A^2 \rangle_{{\rm elprod.}}^{1/2} = (0.639 \pm 0.010)\,$fm,
the {\em axial radius discrepancy} to be discussed in section~\ref{sec:ax}.

\subsection{Data on the induced pseudoscalar form factor}
\label{sec:gpdata}

\begin{figure}[th]
   \epsfysize=7cm
   \centerline{\epsffile{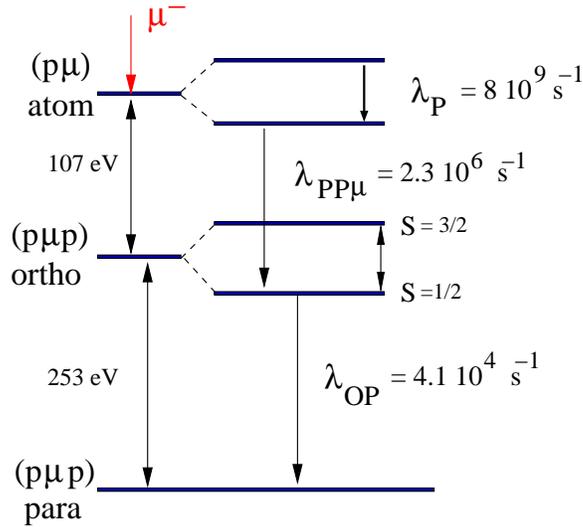}}
   \vspace{.2cm} 
 \caption{\label{fig:omc}
Level scheme for OMC in liquid hydrogen. In the atomic $(p\mu$) state,
one has a triplet/singlet state where the sum of the proton and the muon
spins is 1/0 (upper two levels). In the ortho ($p\mu p$) state, the two
proton spins sum up to one, splitting into two levels with $S=3/2$ and
$S=1/2$ as indicated. Weinberg's mixing suggestion \protect\cite{WeinO}
applies to these two levels.
}
\end{figure}

\begin{figure}[th]
   \epsfysize=7cm
   \centerline{\epsffile{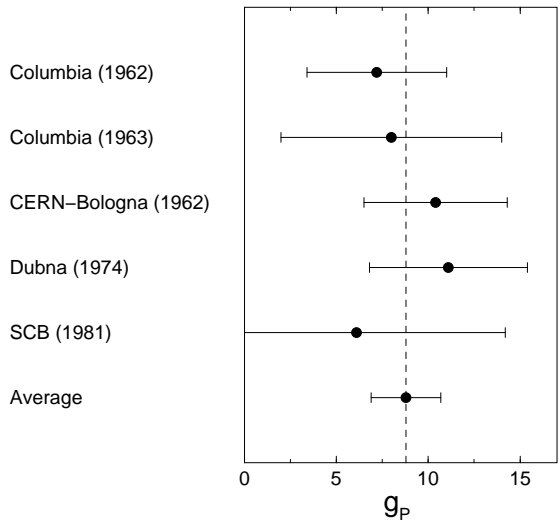}}
   \vspace{.2cm} 
 \caption{\label{fig:gpdat}
 Determination of the pseudoscalar coupling constant
 $g_P$ from (electronics) experiments. The numbers 
 have been rescaled to the present value of $g_A = 1.267$.
 For the measurements on liquid hydrogen, the rate
 $\lambda_{\rm OP} = (4.1 \pm 1.4) \cdot 10^4\,$s$^{-1}$
 is assumed.
 The weighted average is $g_P = 8.79 \pm 1.92$.
}
\end{figure}

\begin{figure}[htb]
   \epsfysize=6cm
   \centerline{\epsffile{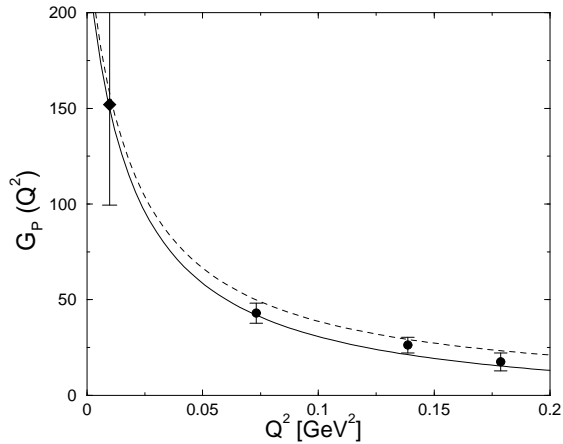}}
   \vspace{.2cm} 
  \parbox{15cm}{\caption{\label{fig:Gptdata}
  The ``world data'' for the induced pseudoscalar form factor
  $G_P(Q^2)$.  The pion electroproduction data
  (filled circles) are from reference~\cite{choi}. Also shown is 
  the world average for ordinary muon capture
  at $Q^2 =0.88M_\mu^2$ (diamond). For orientation, we also show the
  theoretical predictions discussed later.
  Dashed curve: Pion--pole (current algebra) prediction. Solid curve:
  Next--to--leading order chiral perturbation theory prediction. 
  }}
\end{figure}

{\em Ordinary muon capture} (OMC),
\begin{equation}
\label{OMCdef}
\mu^-(l)+p(r)\rightarrow \nu_\mu(l^\prime) +n(r^\prime)~,
\end{equation}
where we have indicated the four--momenta of the various particles,
allows to measure the induced pseudoscalar coupling constant, $g_P$. 
Most experiments so far have used liquid
hydrogen targets, which leads to atomic physics complications because
the $\mu$p atom rapidly forms an p$\mu$p orthomolecule. Consequently,
one has to know the transition rate from the ortho to the para ground
state. Furthermore, there can be mixing between $S=1/2$ and $S=3/2$
components of the ortho p$\mu$p state \cite{WeinO}.
To facilitate the discussion, the various atomic and molecular states
are shown in figure~\ref{fig:omc} together with the pertinent transition
rates (as determined experimentally).
\noindent
The world data on electronics experiments for muon capture in hydrogen
\cite{Columbia1}-\cite{Bardin} are
collected in figure~\ref{fig:gpdat}, these values have been rescaled
to the present day values of the axial--vector coupling constant, the
Fermi constant and the Cabbibo angle. The weighted world average is
\begin{equation}
\label{gpomc}
g_P = 8.79\pm 1.92~,
\end{equation} 
which is in good agreement with theoretical expectations as discussed
below. 
\noindent
At this point it is worth to emphasize that the experimental
value for the ortho--para transition rate $\lambda_{\rm OP} = 
(4.1 \pm 1.4) \cdot 10^4\,$s$^{-1}$~\cite{Bardin} is used to arrive at
the world average. This value disagrees, however, with the theoretical expectation of
reference~\cite{Bakalov}, $\lambda_{\rm OP}^{\rm thy} = 
(7.1 \pm 1.2) \cdot 10^4\,$s$^{-1}$. We also point out that the theoretical 
formula for the liquid capture rate \`a la \cite{Bakalov} involves a time 
integration from zero to
infinity while in the measurement only a finite time interval is
considered, thus a direct comparison is dangerous. A similar
observation has recently been made in reference~\cite{AKMt}.
Stated differently, the theory has to adopt to the experimental
circumstances. We would like to stress here the importance of the new 
proposed experiment at PSI \cite{proposal} which will be done with a
hydrogen gas target and will thus be independent of possible molecular
complications because one will directly measure the atomic singlet
rate. This experiment attempts to measure $g_P$ to
one percent accuracy, which would be a significant improvement
compared to the present situation.
These topics will be taken up again in section~\ref{sec:gp}. 
The induced pseudoscalar coupling constant has also been determined
by muon capture on various nuclei, from light to heavy ones.
A comprehensive review has been given in \cite{Measday}, here we
only report  results for some light nuclei in table~\ref{tab:gp}.
Note that these determinations are mostly in agreement with the
theoretical expectations, but the spread in the results is larger
than given in the table~\cite{Measday}.
In heavier nuclei, there has been much discussion about a possible
quenching of $g_P$ (similar to the quenching of $g_A$), but the
situation can not be considered settled. We do not pursue this topic
here, but refer to the recent reference~\cite{Briancon}.
\begin{table}[htb]
\caption{Pseudoscalar coupling constant determined from OMC
in light nuclei.
}\label{tab:gp}
\begin{indented}
\item[]\begin{tabular}{@{}lcc}
\br
  Nucleus & $g_P$ & Reference \\
\mr
$^3$He (capture to triton)  & $ 8.6 \pm 1.5$ &\protect\cite{Ackerbauer}\\
$^{12}$C (capture to ground state) &  $ 8.3 \pm 2.5$ & \protect\cite{GHMiller}\\
$^{16}$O (capture to $^{16}$N(0$^-$)) &  $ 10.0 \pm 1.2$ & 
\protect\cite{Kane} \protect\cite{Guichon}\\
\br
\end{tabular}
\end{indented}
\end{table}
%
While the momentum transfer in OMC is fixed, {\em radiative
muon capture} (RMC),
\begin{equation}
\label{RMCdef}
\mu^-(l)+p(r)\rightarrow \nu_\mu(l^\prime) +n(r^\prime)+\gamma(k)~,
\end{equation}
has a variable momentum transfer $t$ and one can get up to $t = M_\mu^2$
at the maximum
photon energy of about $k_0 \sim 100\,$MeV, which is quite close to the
pion pole. This amounts approximately to a four times larger
sensitivity to $g_P$ in RMC than OMC. However, this increased
sensitivity is upset by the very small partial branching ratio in hydrogen
$(\sim 10^{-8}$ for photons with $k_0 > 60\,$MeV) and one thus has to
deal with large backgrounds. Precisely for this reason only very 
recently a first measurement of RMC on the proton has been 
published~\cite{TRIUMF1,TRIUMF2}. The resulting
number for $g_P$, which was obtained using a relativistic tree model
including the $\Delta$--isobar~\cite{BF2} to fit the measured
photon spectrum, came out significantly
larger than expected from OMC, 
\begin{equation}
\label{TRIUMFv}
g_P^{\rm RMC} = 12.35 \pm 0.88 \pm 0.38 \simeq 1.4 \, g_P^{\rm OMC}~,
\end{equation}
and thus also about 40\% above all theoretical expectations (see 
section~\ref{sec:Gp}).  It should be noted that in this model the
momentum dependence in $G_P(t)$ is solely given in terms of the pion 
pole and the induced pseudoscalar coupling is obtained as a 
multiplicative factor from direct comparison to the photon spectrum and 
the partial RMC branching ratio (for photon energies larger than 60 MeV). 
We will critically examine this procedure in section~\ref{sec:TRIUMF}.
It was also argued in~\cite{TRIUMF1,TRIUMF2} that the atomic and molecular 
physics related to the binding of the muon in singlet and triplet 
atomic $\mu p$ and ortho and para $p\mu p$ molecular states is 
sufficiently well under control (which is under debate,
see section~\ref{sec:OMC}). This result spurred a lot of theoretical
activity, as discussed later, but here we only wish to point  that it is
a viable possibility
that the discrepancy does not come from the strong interactions but
rather is related to the distribution of the various spin states of the
muonic atoms. It is therefore mandatory to sharpen the theoretical
predictions for the strong as well as the non--strong physics entering
the experimental analysis. In principle, $G_P (t)$ can also be measured
in pion electroproduction, because it enters the longitudinal cross section (in
parallel kinematics)
together with many other effects (see the discussion in section~\ref{sec:pion}).
However, the leading dependence due to the pion pole (and its chiral
corrections) is unique and can be
tested. So far there has only been one experiment~\cite{choi} that
took up the challenge, with the results shown in figure~\ref{fig:Gptdata}.
The observed momentum dependence agrees with theoretical expectations, but
a more refined measurement is certainly called for.  In fact,
at the Mainz Microtron MAMI-B a dedicated experiment has been
proposed to measure the induced pseudoscalar form
factor by means of charged pion electroproduction at low momentum
transfer~\cite{MAMIp}. At this point, it is important to stress that
in the low--energy region, $G_P (t)$ is not truly independent from 
$G_A (t)$, as discussed in detail in section~\ref{sec:Gp}. We conclude that
the induced pseudoscalar form factor is the least well known of all six
electroweak nucleon form factors.


\section{Nucleon axial radius and form factor}
\label{sec:ax}


\subsection{Determination of the axial form factor~I: Neutrino scattering formalism}
\label{sec:nuform}

Neutral and charged current (anti)neutrino scattering off protons
or light nuclei can be used to extract the axial form factor. Since
these topics are well documented in the literature \cite{CLS,Fisch,Kim}, we only
give some basic formulae (sometimes in a highly symbolic notation).
The starting point is the effective Lagrangian of the electroweak
Standard Model for elastic  neutrino--hadron scattering,
such as $\nu_\mu p \to \nu_\mu p$, or quasi--elastic scattering, 
such as $\nu_\mu n \to \mu^- p$ reactions (and similarly for
anti-neutrinos),
\begin{equation}\label{Lnu}
{\cal L}^{\nu h} = -\frac{G_F}{\sqrt{2}}\, \bar{\ell} \, \gamma^\mu
(1+\gamma_5) \nu \, J_\mu^{\rm hadr} + {\rm h.c.}
\end{equation}
with $\ell$ the neutrino/lepton in the final state. We are interested
here in the hadronic current $J_\mu^{\rm hadr}$ which can be written
as
\begin{equation}
J_\mu^{\rm hadr} = \alpha V_\mu^3 + \beta A_\mu^3 + \gamma V_\mu^0
+ \delta A_\mu^0 + \ldots~,
\end{equation}
where the up-- and down--quark contributions are combined to form the
isoscalar and isovector vector ($V_\mu^0, V_\mu^3)$ and axial--vector currents
($A_\mu^0, A_\mu^3)$. The ellipses represent heavy quark ($s,c,b$)
terms. In the Standard Model, one has at tree level
\begin{equation}
\alpha = 1 - 2\sin^2 \theta_W~, \quad \beta = 1~, \quad
\gamma = -\case{2}{3}\sin^2 \theta_W~, \quad \delta = 0~,
\end{equation}
with $\theta_W$ the (weak) mixing angle. It describes the strength of
the mixing of the electromagnetic current with the neutral weak
current and thus only appears in the vector currents. Assuming now
time reversal invariance, isospin invariance, the absence of
second class currents, and neglecting the induced pseudoscalar 
form factor (in some analyses, this form factor is retained 
in its simplified pole form, $G_P(Q^2) \simeq
2mG_A(Q^2)/(Q^2+M_\pi^2)$),  the hadronic current matrix element
between nucleon states is given by
\begin{equation}
\label{wcurrent}
\fl
\langle N(p') \, | J_\mu^{\rm hadr} \, | N(p)\rangle = \bar{u}(p') \,
\left[ \gamma_\mu \, F_1 (Q^2) + \frac{i \sigma_{\mu\nu} q^\nu }{2m} \, 
F_2 (Q^2) +\gamma_\mu \, \gamma_5 \, G_A (Q^2)  \, \right]  \, u(p)~,
\end{equation}
with $Q^2 = -(p' -p)^2 = -q_\mu q^\mu$ the squared momentum transfer. 
Because of vector current conservation, the Dirac and Pauli vector
form factors appearing in equation~(\ref{wcurrent}) are the same
as the ones measured in elastic electron--hadron scattering and can
therefore be considered as known. What is usually assumed in the
analysis of the neutrino scattering experiments is the 
dipole parameterization for the Sachs form factors, $G_E = F_1 +
(Q^2/4m^2)F_2$ and $G_M = F_1 + F_2$. It has the form: $G_E^p = D$,
$G_E^n = 0$, $G_M^{p/n} = \kappa_{p/n} D$, where $D = (1 +
Q^2/M_V^2)^{-2}$, $\kappa_{p/n}$ is the proton/neutron magnetic moment
and $M_V = 0.84\,$GeV. The axial form factor is then extracted
by fitting the $Q^2$-dependence of the (anti)neutrino-nucleon
cross section,
\begin{equation}
{{d}\sigma^{(\nu p, \bar{\nu} p)}\over{d}Q^2}
= \frac{G_F \, m^2}{8 \pi \, E_\nu}\, \left[
A(Q^2) \mp B(Q^2)\,\frac{(s-u)}{m^2} + C(Q^2)\,
\frac{(s-u)^2}{m^4}\, \right]
\end{equation}
in which $G_A (Q^2)$ is contained in the bilinear forms
$A(Q^2)$, $B(Q^2)$ and $C(Q^2)$ of the relevant form factors
and is assumed to be the only unknown quantity
\begin{eqnarray}
\fl
A = \frac{4 \tilde{Q}^2}{m^2} \, \left[ (1+\tau) G_A^2 -  (1-\tau)  F_1^2 +
\tau (1-\tau) F_2^2 + 4\tau F_1 F_2 -\frac{M_\mu^2}{m^2} \left(
  (F_1+F_2)^2 + G_A^2 \right)
\right]~, \nonumber \\
\fl
{B} = 4 \tau \, G_A \, (F_1 + F_2)~, \nonumber \\
\fl
{C} = \case{1}{4} \, \left[ G_A^2 + F_1^2 + \tau \, F_2^2 \right] ~,
\end{eqnarray}
where the plus sign is for neutrinos and the minus sign for
antineutrinos,  and
\begin{equation}
s -u = 4 m E_\nu - \tilde{Q}^2~, \quad \tau = \frac{Q^2}{4m^2}
\end{equation}
with $\tilde{Q}^2 = Q^2 - M_\mu^2$ and $M_\mu$ the muon mass. Of
course, for elastic scattering $\stackrel{(-)}{\nu} p \to 
\stackrel{(-)}{\nu} p$
the terms $\sim M_\mu$ should be dropped.
Due to the smallness of the lepton mass, they are usually also dropped
for the quasi-elastic reactions. Furthermore, $E_\nu$ is the neutrino
energy which is usually determined from the neutrino flux spectrum.
Using the the dipole parameterization
equation~(\ref{dipole}) allows then to determine $M_A$ from a fit to
the measured (anti)neutrino cross section. Here, one assumes that
heavy quark corrections to the axial current can be neglected,
although they are not necessarily small \cite{CWZ,KaMa}. In
particular, the so-called strange quark content of the nucleon
has attracted much attention over the last decade. Since the 
$Q^2$-dependence for such type of contribution is not known, one can simply
modify the dipole ansatz to $G_A (Q^2) = G_{\rm dipole} (Q^2) (1 +
\eta)$, with $\eta$ parameterizing the strange (heavy) quark
contribution. We do not wish to further elaborate on this topic but
rather refer to the references~\cite{ahrens2,nus}.

\subsection{Determination of the axial form factor~II: Electroproduction formalism}
\label{sec:elform}

Consider now pion electroproduction off nucleons,
\begin{equation}\label{elbasic}
\gamma^\star (k^2) + N_1 (p_1) \to \pi^a (q) + N_2 (p_2)~,
\end{equation}
where $\gamma^\star$ denotes the virtual photon with virtuality
$k^2<0$, $N_i (p_i)$ $(i=1,2)$ the initial/final nucleon and
$\pi^a (q)$ the pion with Cartesian isospin $a = (0,+,-)$ and
four--momentum $q_\mu$. We will also use the positive definite
quantity $Q^2 = -k^2$. The pertinent Mandelstam variables are
$s = (p_1+k)^2$, $t = (p_1-p_2)^2$ and $u = (p_1-q)^2$, subject
to the constraint $s+t+u= 2m^2 + M_\pi^2 + k^2$, with $M_\pi$ the 
pion mass. In the Born (one-photon-exchange) approximation, the 
corresponding coincidence cross section can be factorized as \cite{AFF}
\begin{equation}
{{d}\sigma\over{d}E'_{\mathrm{e}}\,
{d}\Omega_{\mathrm{e}}'\,{d}\Omega_\pi^\star} = 
\Gamma_{\mathrm{v}}
{{d}\sigma_{\mathrm{v}}\over
{d}\Omega_\pi^\star}\>{,}
\end{equation}
where $\Gamma_{\mathrm{v}}$ is the virtual photon flux,
$E'_{\mathrm{e}}$, $\Omega_{\mathrm{e}}'$ the energy and
the solid angle of the scattered electron, 
and ${d}\sigma_{\mathrm{v}}/{d}\Omega_\pi^\star$
is the virtual photon cross section in the centre-of-mass frame
of the final $\pi\mathrm{N}$ system, as denoted by the
star.  It can be further decomposed
into transverse, longitudinal and two interference parts,
\begin{equation}
{{d}\sigma_{\mathrm{v}}\over{d}\Omega_\pi^\star}=
{{d}\sigma_{\mathrm{T}}\over{d}\Omega_\pi^\star}+
\epsilon_{\mathrm{L}}^\star\,
{{d}\sigma_{\mathrm{L}}\over{d}\Omega_\pi^\star}+
\sqrt{ 2\,\epsilon_{\mathrm{L}}^\star(1+\epsilon)}\,
{{d}\sigma_{\mathrm{LT}}\over
{d}\Omega_\pi^\star}\,\cos\phi_\pi +
\epsilon {{d}\sigma_{\mathrm{TT}}\over
{d}\Omega_\pi^\star}\,\cos 2\phi_\pi
\end{equation}
with the transverse ($\epsilon$) and longitudinal
($\epsilon_{\mathrm{L}}^\star = -k^2\epsilon/k_0^{\star 2}$) 
polarizations of the virtual photon fixed by the electron kinematics.
In parallel kinematics
($\theta_\pi^\star=\theta_\pi=0^\circ$, with $\theta_\pi$ the
polar angle in the scattering plane as seen in the laboratory system)
the interference parts
vanish. Therefore, at constant four--momentum transfer,
the transverse and the longitudinal cross sections
can be separated using the Rosenbluth method by varying $\epsilon$,
\begin{equation}
{{d}\sigma_{\mathrm{v}}\over{d}\Omega_\pi^\star}=
{{d}\sigma_{\mathrm{T}}\over
{d}\Omega_\pi^\star}
-\epsilon\,{k^2\over k_0^{\star 2}}\,
{{d}\sigma_{\mathrm{L}}\over
{d}\Omega_\pi^\star}~.
\label{eq:rosen}
\end{equation}
(Note that often the photon energy is denoted by $\omega$, however, here
this symbol is entirely reserved for the pion energy).
\noindent
At low energies, 
the connection to theory is most easily made by means of the multipole
expansion. For doing that, one considers the transition current related
to equation~(\ref{elbasic}), which can be decomposed in terms of six
invariant amplitudes (we follow the conventions and notations of
reference~\cite{bkmpr}, see also \cite{berends})
\begin{equation}
\epsilon \cdot T (p_2,s_2;q,a|p_1,s_1;k)
= i\, \bar{u}_2 \,\gamma_5 \, \sum_{i=1}^6 \epsilon \cdot {\cal M}_i
\, A_i (s,u) \, u_1~,
\end{equation}
with $s_i$ the spin index of nucleon $i$. The explicit forms of the
operators ${\cal M}_i$ can be found in \cite{bkmpr} and the $A_i$
are invariant functions that depend on two kinematical variables.
Here, we have chosen the Mandelstam variables $s$ and $u$. The
amplitudes $A_i (s,u)$ have the isospin decomposition
\begin{equation}
A_i (s,u) = A_i^{(+)} (s,u) \, \delta_{a3} + A_i^{(-)} (s,u) \,
\case{1}{2} [\tau_a , \tau_3] +  A_i^{(0)} (s,u) \, \tau_a~.
\end{equation}
In terms of the isospin components, the  physical channels are given by
\begin{eqnarray}
T_\mu (\gamma^\star p \to \pi^+ n) = 
  \sqrt{2} \, [& T_\mu^{(0)} + T_\mu^{(-)}]~, \nonumber \\
T_\mu (\gamma^\star n \to \pi^- p) = 
  \sqrt{2} \, [& T_\mu^{(0)} - T_\mu^{(-)}]~, \nonumber \\
T_\mu (\gamma^\star p \to \pi^0 p) = & T_\mu^{(+)} + T_\mu^{(0)}~, \nonumber \\
T_\mu (\gamma^\star n \to \pi^0 n) = & T_\mu^{(+)} - T_\mu^{(0)}~, 
\end{eqnarray} 
where we have listed the neutral pion channels for completeness. Most of the
following discussions will be focused on the first reaction. For the
discussion of the low--energy theorems that link the axial form factor
to the charged pion production amplitudes, we spell out the corresponding
multipole decomposition of the transition current matrix element at threshold.
In the $\gamma^\star$N centre-of-mass system at threshold, i.e. for $q_\mu =
(M_\pi,0,0,0)$, one can express the current matrix element in terms of the
two S--wave multipole amplitudes, called $E_{0+}$ and $L_{0+}$,
\begin{equation}
\fl
\boldsymbol{\epsilon} \cdot \bi{T} = 4 \pi i (1 + \mu) \, \chi^\dagger_f \, 
\left\{ E_{0+} (\mu, \nu ) + \left[  L_{0+} (\mu, \nu ) - E_{0+} (\mu, \nu )
\right] \, \hat k \cdot \boldsymbol{\epsilon} \, \hat k \cdot 
\boldsymbol{\sigma}  \right\} \, \chi_i~,
\end{equation}
with $\chi_{i,f}$ two--component Pauli--spinors for the nucleon. We have 
introduced the dimensionless quantities
\begin{equation}\label{smallp}
\mu = \frac{M_\pi}{m}~, \quad \nu = \frac{k^2}{m^2}~.
\end{equation}
The multipole $E_{0+}$ characterizes the transverse and $L_{0+}$ the
longitudinal coupling of the virtual photon to the nucleon spin. For an 
explicit expression of these multipoles in terms of the $A_i$ evaluated
at threshold, $s_{\rm thr} = m^2 (1+ \mu)^2$ and $u_{\rm thr} = m^2 ( 1 -
\mu - \mu^2+ \mu \nu)/(1+\mu)$, we refer to \cite{bkmpr}. As it will become
clear in the following paragraphs, the two parameters defined in 
equation~(\ref{smallp}) will play the role of  small expansion parameters in the
QCD analysis of pion electroproduction.

\subsection{Current algebra derivation}
\label{sec:CA}
 
While the methods of current algebra (CA) should be considered outdated
by now, we consider it nevertheless instructive to briefly review the
CA argument which gives the link between charged pion electroproduction
and the nucleon axial form factor. Using standard LSZ formalism, the
transition matrix element can be written as
\begin{equation}
\label{LSZ}
T_\nu^a = - \int d^4x \, {\rm e}^{i q\cdot x} \, ( \square + M_\pi^2)
\langle N_2 | {\cal T} \pi^a (x) V_\nu (0) | N_1 \rangle~,
\end{equation}
with $V_\nu$ the electromagnetic current operator to which the nucleon
couples, $q_\mu$ the pion four--momentum
and ${\cal T}$ is the conventional time--ordering operator.
Inserting now the PCAC relation $\partial^\mu A_\mu^a (x) =
M_\pi^2 F_\pi \pi^a(x)$ and performing a partial integration (note
that while the meaning of the PCAC relation has been understood
since long~\cite{BFGT,Cole}, it does not offer a systematic way of calculating higher
order corrections), one gets
\begin{eqnarray}
T_\nu^a =  {M_\pi^2 -q^2 \over M_\pi^2 F_\pi} \biggl( q^\mu P_{\mu \nu}^a
+ C_\nu^a \biggr)~, \nonumber \\
P_{\mu \nu}^a = i \int d^4x \, {\rm e}^{i q\cdot x} \, 
\langle N_2 | {\cal T} A_\mu^a (x) V_\nu (0) | N_1 \rangle~,
\nonumber \\
C_\nu^a = \int d^4x \, {\rm e}^{i q\cdot x} \, \delta (x_0)\,
\langle N_2 | [A_0^a (x), V_\nu (0)] | N_1 \rangle~.
\end{eqnarray}
Using now the CA relation $[A_0^a (x), V_\nu (0)] = i \epsilon^{a3b}
\delta^3 (\bi{x}) A_\nu^b (0)$, one obtains in the soft pion limit
$q^\mu \to 0$
\begin{eqnarray}\label{Tax}
T_\nu^a =  i \,{\epsilon^{a3b} \over  F_\pi} \,
\langle N_2 (p_2) | A_\nu^b  (0) | N_1 (p_1) \rangle + \mbox{pole terms}~.
\end{eqnarray}
The pertinent nucleon matrix element appearing here was already given
in equation~(\ref{current}), i.e. it contains the axial form factor
$G_A (k^2)$ in terms of the photon virtuality $k^2<0$, since
in the soft--pion limit  $q_\mu  = 0$ the momentum transfer is
$(p_2-p_1)_\mu = k_\mu$.
Because of the $\epsilon$--tensor, the axial form factor
can only appear in charged pion electroproduction. In the chiral limit,
Nambu, Luri\'e and Shrauner~\cite{NLS1,NLS2} have put this in the 
form of a low-energy theorem for the electric dipole amplitude $E_{0+}^{(-)}$,
\begin{equation}
\fl
\quad\quad E_{0+}^{(-)} (M_\pi =0, k^2) = \sqrt{1-{k^2\over 4m^2}} \,
{e \over 8\pi F_\pi}
\, \biggl\{ G_A (k^2) + {g_A k^2 \over 4m^2 -2k^2}G_M^v(k^2)
\biggr\}~,
\end{equation}
with $G_M^v$ the nucleon isovector magnetic form factor. The second term
in this equation stems from the nucleon pole contribution in the chiral limit.
Expanding this to $\Or (k^2)$, one finds
\begin{equation}\label{NLSex}
E_{0+}^{(-)} (M_\pi =0, k^2) = 
{e g_A \over 8\pi F_\pi}
\, \biggl\{ 1 + {k^2 \over 6} \langle r_A^2 \rangle + {k^2 \over 4m}
\biggl(\kappa_v + \frac{1}{2}\biggr) + \Or (k^3) \biggr\}~,
\end{equation}
with $\kappa_v$ the nucleon isovector anomalous magnetic moment.
Of course, this equation is only correct for massless pions with zero
three--momentum. Before the advent of chiral perturbation theory, 
there existed many prescriptions
to get a handle on the corrections due to the pion mass, see
references~\cite{nambu}-\cite{BNR2} For a  textbook 
treatment of this issue, see e.g.\ reference~\cite{AFF}.
This was already addressed in section~\ref{sec:data} in connection
with theoretical uncertainties of the axial form factor extracted
from electroproduction based on various prescriptions to deal
with the corrections beyond the soft--pion limit. 
These uncertainties can and have to be overcome
using the modern methods discussed next. To end this section, we
remark that in principle equation~(\ref{Tax}) should also include
the induced pseudoscalar form factor. However,  in 
soft--pion limit the $G_P (t)$ term is proportional to the photon momentum $k_\mu$,
and when contracted with the photon polarization vector, one obtains the
structure $\boldsymbol{\epsilon} \cdot \boldsymbol{k} \, G_P (t)$. 
Therefore, the induced pseudoscalar
form factor  only contributes to the longitudinal cross section, more precisely
to the longitudinal S--wave multipole $L_{0+}$ (as discussed in section~\ref{sec:pion})
but {\em not} to the transverse multipole $E_{0+}$  which is entirely sensitive to the
axial form factor (radius).

\subsection{QCD analysis of charged pion electroproduction}
\label{sec:Elprod}

\subsubsection{Effective field theory of the Standard Model}
\label{sec:EFT}
In this section, we briefly discuss the effective field theory
of the Standard Model (SM), chiral perturbation theory, without giving
any technical details. For that, we refer the reader to a number
of  reviews \cite{HoGe,Ulf,Pich,Ecker,bkmrev}. 
At energies and momenta much below the scale set by the 
intermediate vector boson masses, the weak interactions are frozen
and the SU(3)$_C \times$SU(2)$_L \times$U(1)$_Y$ Standard Model
simplifies to  SU(3)$_C \times$U(1), i.e QCD in the presence of QED.
These two theories behave very differently at low energies. QED is
characterized by a small coupling constant, $\alpha_{\rm EM} =1/137$,
that means the photons and leptons almost decouple and very precise
perturbative calculations can be performed. On the other hand,
the QCD coupling constant becomes large at low energies, $\alpha_{\rm
  S} = g_{\rm S}^2/4\pi \simeq 1$ and the underlying fields 
(quarks and gluons) are
confined within hadrons. Consequently, to study the strong
interactions, one essentially has to consider
QCD in the presence of external sources embodying the electroweak 
interactions. Apart from lattice gauge theory attempts, which are
plagued by their own conceptual problems, no first 
principle calculations for this theory are possible. However, one
can make use of the {\em symmetries} of QCD and their realizations
to explore in a model--independent way the {\em chiral dynamics} of
QCD. This is based on the observation that the six quark flavors fall
into two distinct classes, the light (u,d,s) and the heavy (c,b,t)
ones (light/heavy compared to the typical hadronic scale of 1 GeV).
Therefore, to a good first approximation one can consider an ideal
world with
\begin{equation}
m_u = m_d = m_s = 0 ~, \quad m_c = m_b = m_t = \infty~,
\end{equation}
i.e. only the light flavors are active and the heavy ones decouple.
This is called the chiral limit of QCD. The corresponding Lagrangian,
${\cal L}_{\rm QCD}^0$, contains only one parameter, $g_{\rm S}$ (or,
by dimensional transmutation, $\Lambda_{\rm QCD}$). ${\cal L}_{\rm
  QCD}^0$ is highly symmetric because the gluon interactions with the
quark triplet $q = (u,d,s)$ are flavor--blind. In particular, one
can decompose the quark fields in  left-- and right--handed
components. These can be independently transformed under
SU(3)$_{L}$ and SU(3)$_{R}$
leaving ${\cal L}_{\rm QCD}^0$ invariant. By Noether's theorem, this
leads to eight conserved vector, $V_\mu^a = \bar{q} \gamma_\mu
(\lambda^a /2)q$, and eight conserved axial--vector,  $A_\mu^a 
= \bar{q} \gamma_\mu \gamma_5 (\lambda^a /2)q$, currents. From these,
one can construct conserved charges,
\begin{equation}
[Q_a^V , H_{\rm QCD}^0 ] = [Q_a^A , H_{\rm QCD}^0 ] \quad (a =1,\ldots,8)~.
\end{equation}  
This is the so--called {\em chiral symmetry of the strong interactions},
which has indeed been observed long before the advent of QCD and led
to current algebra and soft pion techniques (as used in the previous
subsection). However, this symmetry is not necessarily shared by the ground state
or the particle spectrum (hidden or spontaneously broken symmetry). 
So what is the fate of the chiral symmetry in QCD? For any
vector--like gauge theory in the absence of $\theta$--terms, 
one can show that the vacuum is invariant under vector 
transformations~\cite{VaWi}. Concerning the axial transformations,
nature selects the Nambu--Goldstone alternative, i.e the ground state
$|0\rangle$ is not symmetric and the left-- and right--handed worlds
communicate, as indicated by the non--vanishing quark
condensate. Because the axial charges do not annihilate the vacuum,
the spectrum must contain eight pseudoscalar  Goldstone bosons, one
for each broken generator. These particles have zero energy and
three--momentum. As the energy/momentum transfer decreases, the interaction of 
the Goldstone bosons with themselves and matter becomes weak,
paving the way for a perturbative treatment in energies/momenta or
a derivative expansion,
despite the fact that the strong coupling constant is large.
In the real world, the 8 lightest hadrons are indeed
pseudoscalars (the three pions, four kaons and the eta). These
particles are light but not exactly massless simply because the
quark masses are not exactly zero but small. The QCD quark mass term
therefore leads to explicit chiral symmetry breaking, which can be
treated perturbatively. The important observation now is that the
consequences of the broken chiral symmetry can be explored by means
of an effective field theory, called {\em chiral perturbation theory}
(CHPT). In essence, one maps the QCD Lagrangian on an effective Lagrangian
formulated in terms of the asymptotically observed fields, the
Goldstone bosons and matter fields (like nucleons),
\begin{equation}
{\cal L}_{\rm QCD} [q,\bar q , g] \to  {\cal L}_{\rm eff} [ U,
\partial_\mu U, \ldots, {\cal M}, \ldots , N]~,
\end{equation}
where $U$ collects the Goldstone bosons,  the quark mass matrix ${\cal
  M}$ parameterizes the explicit chiral symmetry breaking and matter
fields $N$ can also be included. For a deeper discussion of the
foundations of CHPT, we refer to the seminal papers by
Weinberg~\cite{WeinP}, Gasser and Leutwyler~\cite{GL} 
and Leutwyler~\cite{Leut}. Observables are now expanded in
powers of momenta (energies), generically called $q$, and quark
masses, at a fixed ratio ${\cal M}/q^2$. Stated differently, one
has an underlying power counting~\cite{WeinP} which allows one
to organize any matrix--element $M$ in terms of tree and loop diagrams
for a given chiral dimension, symbolically
\begin{equation}\label{power}
M = q^\nu \, f \left( \frac{q}{\mu} , g_i \right)~,
\end{equation}
where $q$ is a small momentum or meson mass, $f$ a function of order
one, $\mu$ a regularization scale and $g_i$ a collection of coupling 
constants. Because of chiral symmetry, the counting index (chiral
dimension) $\nu$ is bounded from below. One can show that n--loop
graphs are suppressed by powers of $q^{2n}$, thus the leading
contribution to any given process stems from tree graphs with the
lowest order insertions (which is equivalent to current algebra).
To a given order, the corresponding  effective Lagrangian must contain
all terms compatible with chiral, gauge and discrete symmetries.
Beyond leading order, such local interactions are accompanied by
unknown coupling constants (the $g_i$ in equation~(\ref{power})),
also called low--energy constants (LECs). While these can in principle
be calculated from QCD, $g_i = g_i (\Lambda_{\rm QCD},m_c,m_b,m_t)$,
in practice they must be determined by a fit to some data or estimated
using some model. Let us illustrate this for the pion--nucleon system
chirally coupled to external sources like e.g. electroweak gauge
bosons. A complete one--loop calculation is based on the effective
Lagrangian
\begin{equation}
{\cal L}_{\rm eff} = {\cal L}_{\pi N}^{(1)} + {\cal L}_{\pi N}^{(2)}
+ {\cal L}_{\pi N}^{(3)} +{\cal L}_{\pi N}^{(4)}~,
\end{equation} 
from which one calculates tree graphs with insertions from all terms
$ {\cal L}_{\pi N}^{(1,2,3,4)}$
and loop diagrams with at most one insertion from the dimension two
operators collected in ${\cal L}_{\pi N}^{(2)}$. The lowest order
effective Lagrangian contains $g_A$ and $m$ (note that the nucleon mass might be
transformed into higher order $1/m$ corrections~\cite{jm}). At order two, three
and four one has 7, 23 and 118 LECs, respectively. While these numbers
appear large, it should be stressed that many of the dimension four operators 
contribute only to very exotic processes, like three or four pion
production induced by photons or pions. Moreover, for a given
process, it often happens that some of these operators appear in
certain linear combinations and others simply amount to a quark mass renormalization
of the corresponding dimension two operator.
For example, in the case of elastic pion--nucleon scattering, one has
4, 5, and 4 independent LECs (or combinations) thereof at second, third
and fourth order, respectively, which is a much smaller number than
the total number of
independent terms. Consequently, the often cited folklore that CHPT
becomes useless beyond a certain order because the number of LECs increases
drastically is not really correct. It is also important to stress that
the values of the dimension two (and of some dimension three) LECs can
be understood in terms of (low--lying)  t--channel meson resonance and s--channel
baryon resonance excitations~\cite{bkmLEC} (the so--called resonance
saturation). For further details, we refer the
reader to references~\cite{bkmrev,ulfrev}.

\subsubsection{Low--energy theorems for pion electroproduction}
\label{sec:LET}

\begin{figure}[htb]
   \epsfysize=6cm
   \centerline{\epsffile{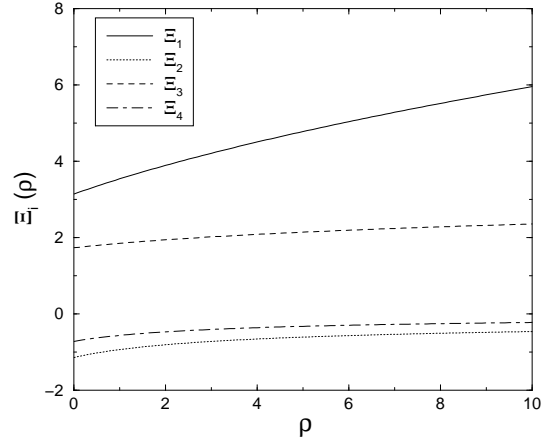}}
   \vspace{.2cm} 
   \centerline{\parbox{15cm}{\caption{\label{fig2}
   The functions $\Xi_i (\rho)$ ($i=1,2,3,4)$ as defined in 
   equations~(\ref{xi}).
  }}}
\end{figure}

We now wish to apply the machinery of baryon CHPT to the case of pion
electroproduction and derive low--energy theorems (LETs) . Before doing that,
a few clarifying remarks about the meaning of such LETs are in order.
Following the seminal work of Nambu \etal \cite{NLS1,NLS2}, 
pion electroproduction LETs have been discussed in the sixties, often
using by now outdated methods, see~\cite{FFR,RL,AG,Ad}. As already
stressed, only using the methods of CHPT one can formulate
model--independent statements beyond leading order, provided one
is able to fix all LECs from data. The modern meaning of LETs is
discussed in broad detail in reference~\cite{EM}.   To third order
in the chiral expansion, the various isospin components of the
transverse and the longitudinal S--wave multipoles can be written
as functions of the two small parameters $\mu$ and $\nu$, cf.
equation~(\ref{smallp}), as (for details, see \cite{bkmpr})
\begin{eqnarray}
\fl
E_{0+}^{(+)} (\mu , \nu) = {e g_{\pi N}  \over 32\pi m }\,
  \biggl\{ -2 \mu + \mu^2 (3+\kappa_v) - \nu (1+ \kappa_v)
+ {\mu^2 m^2 \over
  4\pi^2 F_\pi^2} \Xi_1(-\nu\mu^{-2}) \biggr\} + \Or(q^3)~, \\
\fl
L_{0+}^{(+)} (\mu , \nu)  =  E_{0+}^{(+)} (\mu , \nu) 
+ {e g_{\pi N}  \over 32\pi m }\,
(\mu^2 -\nu) \biggl\{ \kappa_v 
+  {m^2\over 4\pi^2 F_\pi^2}\Xi_2(-\nu\mu^{-2}) \biggr\} + \Or(q^3)~, \\
\fl
E_{0+}^{(0)} (\mu , \nu) = {e g_{\pi N}  \over 32\pi m }\,
  \bigl\{ -2 \mu + \mu^2 (3+\kappa_s) - \nu (1+ \kappa_s)
  \bigr\} + \Or(q^3)~, \\
\fl
L_{0+}^{(0)} (\mu , \nu)  =  E_{0+}^{(0)} (\mu , \nu) 
+ {e g_{\pi N}  \over 32\pi m }\,
(\mu^2 -\nu) \, \kappa_s + \Or(q^3)~, \\
\fl
E_{0+}^{(-)} (\mu , \nu) = {e g_{\pi N}  \over 8\pi m }\,
  \biggl\{ 1 - \mu +C\mu^2 
+ \nu \biggl(  {\kappa_v \over 4} + {1\over
  8} + {m^2\over 6} \langle r_A^2\rangle \biggr)
+ {\mu^2 m^2 \over
  8\pi^2 F_\pi^2} \Xi_3(-\nu\mu^{-2}) \biggr\} \nonumber \\
\lo+ \Or(q^3)~,  \\
\fl
L_{0+}^{(-)} (\mu , \nu)  =  E_{0+}^{(-)} (\mu , \nu) 
+ {e g_{\pi N}
  \over 8\pi m }\,
(\mu^2 -\nu) \biggl\{ {\kappa_v \over 4} + {m^2\over 6} \langle r_A^2
\rangle 
+ {\sqrt{(2+\mu)^2 -\nu} \over 2(1+\mu)^{3/2} (\nu - 2\mu^2
  -\mu^3)} \nonumber \\
\lo+ \biggl( {1\over \nu -2\mu^2} - {1\over \nu} \biggr)
(F_\pi^V (m^2\nu) -1)
+  {m^2\over 8\pi^2 F_\pi^2}\Xi_4(-\nu\mu^{-2}) 
\biggr\} + \Or(q^3)~, \label{L0pm}
\end{eqnarray}
with $\kappa_v = \kappa_p - \kappa_n = 3.71$ the isovector
and $\kappa_s = \kappa_p + \kappa_n =-0.12$ isoscalar anomalous magnetic moment
of the nucleon, $g_{\pi N}$ the strong pion--nucleon coupling constant
(which can be used instead of $g_A$ by virtue of the
Goldberger-Treiman relation) and $F_\pi^V(k^2)$ is vector (charge)
form factor of the pion. The functions $\Xi_j (-\nu / \mu^2)$,
$(j=1,2,3,4)$ are given by
\begin{eqnarray}
\fl
\Xi_1 (\rho) = {\rho \over 1+\rho} + {(2+\rho)^2 \over 2(1+\rho)^{3/2}}
\arccos {-\rho \over 2+\rho}~, \nonumber \\
\fl
\Xi_2 (\rho) = {2-\rho \over (1+\rho)^2} - {(2+\rho)^2-2 \over 2(1+\rho)^{5/2}}
\arccos {-\rho \over 2+\rho}~,  \nonumber \\
\fl
\Xi_3 (\rho) = \sqrt{1+{4\over \rho}} \ln \biggl(\sqrt{1+{\rho
  \over 4}} + {\sqrt{\rho} \over 2} 
   + 2 \int_0^1 \sqrt{(1-x) [
  1 + x(1+\rho )] } \nonumber \\
\lo{\times} \arctan {x\over \sqrt{(1-x)[1+x(1+\rho)]} }~,  \nonumber\\
\fl {\Xi_4} (\rho) =  \int_0^1 dx {x(1-2x)\over
  \sqrt{(1-x)[1+x(1+\rho)]}} \,
  \arctan {x\over \sqrt{(1-x)[1+x(1+\rho)]
  }}~. \label{xi}
\end{eqnarray}
These functions are shown in figure~\ref{fig2} for $0<\rho <10$. They
exhibit a very smooth behavior. The influence of isospin breaking,
i.e. the charged to neutral pion mass difference in the loops, has been
discussed in reference~\cite{bkmpr}, here it suffices to say that the
$\rho$--dependence is only very little affected by such effects.
\noindent Note that the LETs given above do not contain the full
electromagnetic form factors of the nucleon. This is due to the power
counting of CHPT, from which one concludes that at third order one
is only sensitive to the normalization of the magnetic form factors,
i.e. the respective anomalous magnetic moments. At fourth order, one
expects sensitivity to the electric charge radii of the proton and the
neutron. Of particular interest are the LETs for the $(-)$ amplitudes 
because of the connection to the nucleon axial form factor, as will be
discussed in the next paragraph. Here, we just mention that the
constant $C$ appearing in the expression for $E_{0+}^{(-)}$ can be
expressed in terms of parameters that can be fixed in pion
photoproduction, see~\cite{bkmpr}. Its explicit form is, however,
not needed in what follows.

\subsubsection{Nucleon axial radius}
\label{sec:ra}

\begin{figure}[htb]
\begin{center}
\hspace{0.8cm}
\epsfysize=2.8cm
\epsffile{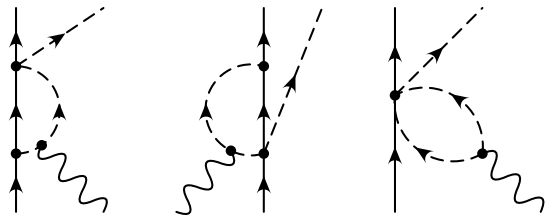}
\vspace{0.3cm}
\caption{One--loop  diagrams that lead to the axial radius correction. 
Crossed partners are not shown. The solid, dashed and wiggly lines 
denote nucleons, pions and photons, in order.\label{fig:diaax}}
\end{center}
\end{figure}

From the LETs just given, we can now derive the relation between
the electroproduction amplitude and the nucleon axial radius,
following reference~\cite{bkmax}. To separate the axial radius, 
one should consider the slope of the transverse multipole. 
For doing that, one has to expand the function $\Xi_{3}
(-\nu\mu^{-2})$ in powers of $k^2 = \nu m^2$ and pick up all 
terms proportional to $\nu$. However, to make the argument even
clearer, we give one intermediate step. In fact, the direct
diagrammatic evaluation of the electric dipole amplitude to third
order in the chiral expansion includes the one--loop diagrams
shown in figure~\ref{fig:diaax}. 
\noindent
These lead to an additional
contribution to $E_{0+}^{(-)}$ at order $k^2$ that can not 
be obtained using current algebra methods,
\begin{eqnarray}\label{dis3i}
\fl
E_{0+}^{(-)} (M_\pi \neq 0, k^2) & = \frac{e g_A}{8 \pi F_\pi}
\left\{ 1 + \frac{k^2}{6} \langle r_A^2 \rangle + \frac{k^2}{4m^2}
\left( \kappa_v + \frac{1}{2} \right) - \mu + C\mu^2 \right. \nonumber \\
\fl
&\left. + \frac{\mu^2 m^2}{8 \pi^2 F_\pi^2} \int_0^1 dx \int_0^x dy \,
\ln \left[ 1-x^2+y^2 + \frac{\nu}{\mu^2} x (x-1) \right] \right\}
+ \Or(q^3)~.
\end{eqnarray}
Note that in a third order calculation, the electric dipole amplitude
is only given to second order in small momenta. The constant $C$ in
(\ref{dis3i}) subsumes various $k^2$--independent loop corrections
together with contact terms and kinematical corrections of order
$1/m^2$. In the chiral limit $\mu =0$, one recovers of course the
LET of Nambu \etal \cite{NLS1,NLS2}, because the last term
of equation~(\ref{dis3i}) vanishes identically at zero pion
mass in the physical region $k^2<0$, compare also equation~(\ref{NLSex}) and
the discussion below. Matters are, however, different for
non--vanishing pion mass, $\mu \neq 0$. Expanding the ln-term in
(\ref{dis3i}) in powers of $k^2$, the coefficient of the $k^2$--term
in equation~(\ref{NLSex}) receives an extra contribution,
\begin{equation}
 1 + {k^2 \over 6} \langle r_A^2 \rangle + {k^2 \over 4m}
\biggl(\kappa_v + \frac{1}{2}\biggr) + \frac{k^2}{128F_\pi^2}
\biggl( 1 - \frac{12}{\pi^2} \biggr) ~.
\end{equation}
The last term in this equation constitutes a model--independent
contribution at order $k^2$ not modified by higher loop or 
contact terms, which are suppressed by powers of $\mu$.
Formally, it appears because one cannot interchange the limits
$M_\pi \to 0$ (chiral limit) and $k^2 \to 0$ (photon point).
Therefore, what was believed to be the axial radius determined 
in pion electroproduction was nothing but the modified radius
\begin{equation}
\langle \tilde{r}_A^2 \rangle = \langle {r}_A^2 \rangle +
\frac{3}{64 F_\pi^2} \biggl( 1 - \frac{12}{\pi^2} \biggr) ~.
\end{equation}
The numerical value of this additional term is 
\begin{equation}
\Delta \langle {r}_A^2 \rangle = 
\langle \tilde{r}_A^2 \rangle - \langle {r}_A^2 \rangle
= -0.0456\,{\rm fm}^2~,
\end{equation}
which translates into a shift of the axial mass of
\begin{equation}
\Delta M_A = 0.055~{\rm GeV}~,
\end{equation}
which is in agreement with the empirical value given in
equation~(\ref{axdis}), $\Delta M_A=(0.043\pm 0.026)\,\mathrm{GeV}$. 
This agreement can be considered as one of the premier successes of
baryon chiral perturbation theory. 

\medskip \noindent
Since the lowest (third) order result for the axial radius discrepancy
is quite small, one should investigate higher order corrections in $\mu$
to find out how reliable the leading order result is. This has been
reported in \cite{schmidt} and we essentially follow the arguments given
there. To order $q^4$, which gives the pertinent 
next-to-next-to-leading order corrections, one
has to consider one loop graphs with exactly one insertion from the dimension
two $\pi$N Lagrangian and counterterms. The LECs related to these will
be estimated from resonance exchange contributions, here $\rho$ and
$\Delta$ exchanges. Since we are only interested in the transverse
part of the transition matrix element for $\gamma^\star p \to \pi^+ n$
at threshold, we introduce an auxiliary quantity $E(k^2)$, that is
proportional to the transverse threshold S--wave multipole,
\begin{equation}
\boldsymbol{T} \cdot \boldsymbol{\epsilon} = \frac{ie g_A}{\sqrt{2} F_\pi} \,
\boldsymbol{\sigma} \cdot \boldsymbol{\epsilon} \, E(k^2) 
= 4\pi i (1+\mu) \, \boldsymbol{\sigma} \cdot \boldsymbol{\epsilon} \,
E_{0+, {\rm thr}}^{\pi^+ n}~.
\end{equation}
In the chiral limit, the corresponding current algebra result becomes
exact and gives, when expanded in $k^2$
\begin{equation}
E(k^2) = 1 + \frac{k^2}{6} \langle \krig{r}_A^2 \rangle -
\frac{k^2}{2m} \left( \krig{\kappa}_n + \frac{1}{4} \right)
+ \Or(k^4)~,
\end{equation}
with
\begin{equation}
 \langle \krig{r}_A^2 \rangle = \langle {r}_A^2 \rangle + \Or(\mu^2)~,
\quad \krig{\kappa}_n = \kappa_n - \frac{g_A^2 m M_\pi}{8\pi F_\pi^2}
 + \Or(\mu^2)~,
\end{equation}  
the nucleon axial radius and the neutron magnetic moment in the chiral
limit. Here the superscript '$\krig{}$' denotes quantities in the chiral limit,
${\cal Q}= \krig{{\cal Q}} [ 1 + {\cal O}(m_{u,d})]$. 
Note that the axial radius (like the isocalar electromagnetic
radius) is analytic in the quark masses. We now want to calculate all
tree and loop graphs up to order $q^4$ which contribute to the slope
$E'(0) = \partial E(0) / dk^2$ and are proportional to $\mu^0$ and/or
$\mu$. The quantity $E'(0)$ is the sum of the desired squared axial 
radius and a host of other terms.  Among these are contributions from
the Born graphs including electromagnetic form factors. Since these are
already contained in the standard analysis of the pion electroproduction
data, such terms need not be considered further. The discrepancy subsumes
all loop and counterterm effects which go beyond the form factors. After
some lengthy calculation, one arrives at
\begin{eqnarray}\label{dis4}
\fl
\langle \tilde{r}_A^2 \rangle - \langle {r}_A^2 \rangle  = &
\frac{3}{64 F_\pi^2} \left( 1 - \frac{12}{\pi^2} \right) + 
\frac{3M_\pi}{64 mF_\pi^2} + \frac{3c^+ (\pi-4) M_\pi}{32\pi F_\pi^2}
\nonumber \\
& + \frac{3g_A^2 M_\pi}{8 \pi^2 m F_\pi^2} \left( \ln \frac{M_\pi}{\lambda}
- \frac{\pi^2}{16} + \frac{7\pi}{12} - \frac{1}{4} \right)
+ 6E'_\rho (0) +  6E'_\Delta (0)~.
\end{eqnarray}
The first term in (\ref{dis4}) is the leading order result already
given before.  The combination of LECs, $c^+ = -8c_1 + 4c_2 + 4c_3 - g_A^2/2m$,
can be related to the isospin--even $\pi$N scattering length $a^+_{0+}$
via $c^+ = 8\pi F_\pi^2 a^+_{0+}/M_\pi^2$ (which is correct to the
order we are working here). The last two terms in (\ref{dis4}) represent
the counterterm contributions at order $q^4$, which have been identified
with $\rho (770)$ and $\Delta (1232)$ exchange contributions,
\begin{eqnarray}\label{ct4}
6E'_\rho (0) =  & -\frac{3(1+ \kappa_\rho) M_\pi}{16 \pi^2 g_A m F_\pi^2}
~, \nonumber \\
 6E'_\Delta (0) = & \frac{\kappa^\star M_\pi}{\sqrt{2} m^2 m_\Delta^2}
\biggl[ {m_\Delta^2 - m_\Delta m + m^2 \over m_\Delta -m } 
\nonumber \\
&  - 2m (Y+Z+2YZ) - 2m_\Delta (Y+Z+4YZ) \biggr]~,
\end{eqnarray}
with $\kappa_\rho \simeq 6$ the tensor--to--vector ratio of the 
$\rho NN$ couplings and $\kappa^\star = g_1 \simeq 5$ is the 
$\gamma \Delta N$ coupling constant. The so--called off--shell
parameters $Y$ and $Z$ (which are nothing but higher order LECs
in an effective field theory including the delta) have been
constrained in reference~\cite{schmidt} from the delta contribution
to the proton's magnetic polarizability, $Y =-0.12$, and from 
the $a_{33}$ $\pi$N scattering volume, $Z=-0.3$.
Putting all pieces together, one finds for the axial radius
discrepancy,
\begin{equation}\label{valdis4}
\langle \tilde{r}_A^2 \rangle - \langle {r}_A^2 \rangle  = 
(-4.6 + 0.5) \cdot 10^{-2} \, {\rm fm}^2~.
\end{equation}
Note that there are large individual terms in the loop and the
counterterm contributions which cancel each other. Also, this
correction is fairly sensitive to the badly known isoscalar
$\pi$N scattering length, so it is fair to assign an uncertainty to
this correction which is as large as it central value given
in equation~(\ref{valdis4}). However, the only meaningful quantity
is the total sum of all terms of order $q^4$, which can also be
considered verified by the MAMI experiment~\cite{MAMI}. Taken that
experiment face value, one could turn the argument around and give
a bound on $a^+_{0+}$. We refrain from doing that here but stress
again the intricate interplay of seemingly unrelated processes,
linked by the chiral structure of QCD. It should also be noted that
the momentum transfers in the MAMI experiment were too high to allow
for a direct and safe extraction of $E_{0+}^{(-)}$, therefore the
data were analyzed in the framework of an effective Lagrangian model
with the electromagnetic nucleon form factors, the pion vector form
factor and the axial nucleon form factor at the appropriate 
vertices~\cite{A1,DT}. A further  measurement at lower energy and lower
momentum transfer should therefore be studied for feasibility.

\subsubsection{Pion charge radius}
\label{sec:pion}

\noindent The charge (vector) radius of the pion is a fundamental quantity in
hadron physics. It can essentially be determined in two ways. One method is
pion scattering off electrons (or electron--positron
annihilation into pion pairs), this leads a pion root--mean--square
(rms) radius of~\cite{Amen}
\begin{equation}
\langle r^2_\pi \rangle_V^{1/2} = (0.663 \pm 0.006)~{\rm fm}~,
\end{equation}
if one insists on the correct normalization of the pion charge (vector)
form factor, $F_\pi^V (0) =1$ (in units of the elementary charge $e$).
A more recent determination of the pion vector radius from low--momentum
space-- and time--like form factor data based on a very precise two--loop 
chiral perturbation theory representation~\cite{BCT} gives (two--loop
contributions to the pion form factors were first considered in 
reference~\cite{GM})
\begin{equation}
\langle r^2_\pi \rangle_V^{1/2} = (0.661 \pm 0.012)~{\rm fm}~, 
\end{equation}
consistent with the value given above. Electromagnetic corrections
considered in reference~\cite{KM} reduce this value by about one
percent. This number can be understood semi--quantitatively in a naive vector
meson dominance picture, $\langle r^2_\pi \rangle_V^{1/2} =
\sqrt{6}/M_\rho \simeq 0.63\,$fm. The second method is based on
charged pion electroproduction, $\gamma^\star p \to \pi^+ n$.
The unpolarized
cross section in parallel kinematics decomposes into a transversal
and a longitudinal piece, as detailed in section~\ref{sec:LET}.
While the former is sensitive 
to the the nucleon axial radius, the latter is quite sensitive to the 
pion form factor, i.e. to the pion radius for small momentum transfer. 
A recent  measurement at the Mainz Microtron MAMI--II led to a pion radius
of~\cite{MAMI}
\begin{equation}\label{MAMIrad}
\langle r^2_\pi \rangle_V^{1/2} = (0.74 \pm 0.03)~{\rm fm}~,
\end{equation}
which is a sizeably larger value than the one obtained from $\pi e$
scattering. It was hinted in reference~\cite{MAMI} that their larger
value for the pion radius might be due to the inevitable
model--dependence based on the Born term approach to extract the
pion radius. It was also stated that there might be an additional
correction obtainable form chiral perturbation theory as it is the
case for the nucleon axial radius discussed before.
Indeed,  there is such   a similar kind of correction for the pion 
radius~\cite{BKMpi}.
This new term modifies the momentum dependence of the longitudinal
S--wave amplitude $L_{0+}^{(-)}$ and leads one to expect an
even larger pion charge radius than the one given in
equation~(\ref{MAMIrad}). It is conceivable that higher order corrections
yet to be calculated or contributions from higher multipoles
will completely resolve the discrepancy between the pion radius
determined from $\pi e$ scattering on one side and from charged pion
electroproduction on the other. Let us be more specific now.
As shown in equation~(\ref{L0pm}), the pion vector form factor 
$F_\pi^V (k^2)$ appears in the expression for the longitudinal 
multipole.  The pion form factor has the following low--energy expansion,
\begin{equation}
F_\pi^V (k^2) = 1 + {1\over 6} \langle r^2_\pi \rangle_V \, k^2 +
{\cal O}(k^4)~.
\end{equation}
Consequently, to separate the pion radius, one should consider the
slope of the longitudinal multipole. For doing that, one has to expand
the function $\Xi_{4} (-\nu\mu^{-2})$ in powers of $k^2 = \nu m^2$
and pick up all terms proportional to $\nu$. This gives:
\begin{eqnarray}\label{L0}
\fl
{\partial L_{0+}^{(-)} \over \partial k^2} \biggl|_{k^2 = 0} =
{e g_{\pi N} \over 32 \pi m} \biggl\{ {1 \over M_\pi^2} - {1\over
  mM_\pi} + {13\over 8m^2} 
+ {1\over 3} \langle r_\pi^2 \rangle_V &+&
{1\over 32 F_\pi^2} \biggl( {16\over \pi^2} -1 \biggr)
+ {\cal O} (M_\pi) \biggr\}~.
\end{eqnarray}
The first four terms are standard~\cite{scherer}, they comprise the 
conventional dependence on the pion vector radius, recoil effects and the
dominant chiral limit behavior of the slope of the longitudinal
multipole. The strong $1/M_\pi^2$ chiral singularity stems from the
$k^2$--derivative of the pion--pole term which appears already at leading order
in the chiral expansion (this is the leading contribution from the
{\em induced pseudoscalar form factor}).  
The last term in equation~(\ref{L0}) originates from the so--called
triangle and tadpole (with three pions coupling to the nucleon at one
point) diagrams which are known to play a prominent role in pion photo--
and electroproduction (see figure~\ref{fig:diaax}). 
The formal reason for the appearance of 
this new, model--independent contribution at order $k^2$ is that one cannot
interchange the order of taking the derivative at $k^2 =0$ and the chiral limit
$M_\pi \to 0$. 
Consequently, all determinations of the pion radius from
electroproduction (based on tree-level amplitudes including nucleon and pion
form factors) have ``measured'' the modified radius,
\begin{equation}\label{modrad}
 \langle \widetilde{r}^2_\pi \rangle_V =  \langle r^2_\pi \rangle_V
+ {3\over 32 F_\pi^2} \biggl( {16\over \pi^2} -1 \biggr)~.
\end{equation}
The novel term on the right--hand-side of equation~(\ref{modrad}) amounts
to 0.266~fm$^2$, a bit more than half of the squared pion rms radius,
$\langle r_\pi^2 \rangle_V \simeq 0.44\,$fm$^2$. Therefore, from the
longitudinal multipole alone, one expects to find a larger
pion radius if one analyses pion electroproduction based on Born
terms, 
\begin{equation}
 \langle \widetilde{r}^2_\pi \rangle_V = (0.44+0.26)~{\rm fm}^2 =
(0.83~{\rm fm})^2~,
\end{equation}
which is even larger than the result of the Mainz analysis, 
cf. equation~(\ref{MAMIrad}). We point out, however, that the 
contribution of the pion radius to the derivative of the longitudinal
multipole is a factor of ten smaller than the one from the first
three terms in the curly brackets in  equation~(\ref{L0}). Therefore,
a fourth order analysis is certainly needed to further quantify
the ``pion radius discrepancy''. Furthermore, the pion form factor
contribution to the longitudinal cross section is also present in higher
multipoles. In fact, it is known that the convergence of the multipole
series for the pion pole term is slow. One should therefore also 
investigate such effects for these higher multipoles or directly compare the
predictions of complete one--loop calculation with the data of the
longitudinal electroproduction cross section.   For the purpose of
demonstrating the significance of chiral loop effects the $k^2$-slope
is, however, best suited.  What has been shown here is 
that as in the case of the nucleon axial 
mean square radius, the pion loops, which are a unique consequence of the
chiral symmetry of QCD, modify the naive Born term analysis and should be
taken into account. 

\section{Induced pseudoscalar form factor and coupling constant}
\label{sec:gp}

\subsection{QCD analysis of the induced pseudoscalar form factor}
\label{sec:Gp}

Here, we show that one can give an accurate prediction for the induced
pseudoscalar form factor $G_P (t)$  and coupling constant $g_P$ in terms of
well--known physical parameters, following reference~\cite{bkmgp} (for
a similar analysis, see \cite{FLMS}). For doing that, one exploits the chiral Ward
identity of two--flavor QCD,
\begin{equation}
\partial^\mu \, \left( \bar{q} \, \gamma_\mu \gamma_5 \, \frac{\tau^a}{2} q \right) =
\hat{m} \, \bar{q} \, i \gamma_5 \, \tau^a \, q
\label{e3}
\end{equation}
with $\hat m$ the average light quark mass (note that
isospin--breaking effects can be safely neglected here).
Sandwiching equation~(\ref{e3}) between nucleon states, one obtains \cite{GSS}
\begin{equation}
mG_A(t) + \frac{t}{4 m} G_P(t) = 2 \hat{m} \, B \, \krig{m} \,
 \krig{g}_A \frac{1 + h(t)}{M_\pi^2 - t}
\label{e4}
\end{equation}
The pion pole in equation~(\ref{e4}) originates
from the direct coupling of the pseudoscalar density  to the pion,
$\langle 0|\bar{q}
i \gamma_5 \tau^a q | \pi^b \rangle = \delta^{ab} G_\pi$ \cite{GL}. The residue at the
pion pole $t = M_\pi^2$ is \cite{GSS}
\begin{equation}
\hat{m} \, G_\pi \, g_{\pi N} = g_{\pi N} \, F_\pi \, M_\pi^2
\label{e5}
\end{equation}
with $g_{\pi N}$ the strong pion--nucleon coupling constant.
The chiral expansion of the axial form factor and of the function $h(t)$
to order $q^4$ takes the form
\begin{equation}
G_A (t) =  g_A \left( 1 + \frac{\langle r_A^2 \rangle}{6} t \right)~, \quad
h(t) =  {\rm const}  - \frac{2 \bar{d}_{18}}{g_A} t
\label{e6}
\end{equation}
with $\bar{d}_{18}$ a low--energy constant (we use the notation of reference~\cite{FMS})
and one does not need to specify
the constant since it is not needed explicitly in the following.
The reason for the linear dependence in equation~(\ref{e6}) is the following.
The corresponding form factors $G_A (t)$ and
$h(t)$ have a cut starting at $t = (3 M_\pi)^2$ which in the chiral expansion
first shows up at two--loop order ${\cal O}(q^5)$.
Therefore, the contribution to order $q^4$ must be polynomial in $t$.
Furthermore,  from chiral counting it follows that the
$t^2$-terms are related to order $q^5$ of the full matrix--elements. Putting
pieces together, we arrive at
\begin{equation}
m \, g_A + m \, g_A \frac{\langle r_A^2 \rangle}{6} t + \frac{t}{4m} G_p(t) =
\frac{g_{\pi N} F_\pi}{M_\pi^2-t} t + g_{\pi N} F_\pi + \frac{2 \bar{d}_{18} M_\pi^2
g_{\pi N}}{g_A}
\label{e7}
\end{equation}
using $2 \hat{m} B \krig{g}_A \krig{m} = M_\pi^2 ( g_{\pi N}
F_\pi + {\cal O}(M_\pi^2) )$. At $t=0$, equation~(\ref{e7}) reduces to the
Goldberger--Treiman discrepancy (GTD) \cite{GSS,BKKM}
\begin{equation}
 g_A \, m = g_{\pi N} \,F_\pi
\, \biggl( 1+ \frac{2 \bar{d}_{18}}{g_A} \, M_\pi^2 \biggr)~,
\label{e8}
\end{equation}
which also clarifies the meaning of the low--energy constant
$\bar{d}_{18}$. We will return to a more detailed discussion of the
GTD in section~\ref{sec:GTR}.
Finally, $G_P (t)$ can be isolated from equation~(\ref{e7}),
\begin{equation}
G_P (t) = \frac{4 m g_{\pi N} F_\pi}{M_\pi^2 - t} \, - \frac{2}{3} \, g_A \,
m^2 \, \langle r_A^2 \rangle + \Or(t, M_\pi^2)~.
\label{e9}
\end{equation}
A few remarks are in order. First, notice that only physical and
well--determined parameters enter in this formula. Second, while the first
term on the right--hand--side  is of order $q^{-2}$, the second
one is ${\cal O}(q^0)$ and the corrections not calculated are of order
$q^2$. Third, the first term is the celebrated pion pole term, which
stems from the direct coupling of the pion field to the
pseudoscalar source. It clearly dominates this form factor for
momentum transfers of the order of a few pion masses. Also, because of
the pion pole, the structure of this form factor is not of the common
form, i.e. its low--energy expansion is not a polynom characterized
by some normalization and a radius. Because of the Ward identities, it
is linked to the axial form factor as witnessed by the appearance of
the axial radius in equation~(\ref{e9}).
For the pseudoscalar coupling $g_P$, this leads to
\begin{equation}
g_P  = \frac{2 M_\mu g_{\pi N} F_\pi}{M_\pi^2 + 0.88M_\mu^2} \,
 - \frac{1}{3} \, g_A \, M_\mu \, m \, \langle r_A^2 \rangle~.
\label{e10}
\end{equation}
Indeed, this relation has been derived long time ago by Adler and
Dothan~\cite{AD}
and by 
Wolfenstein \cite{WOL}. Wolfenstein used a once--subtracted dispersion relation for the
right--hand--side of equation~(\ref{e4}) (weak PCAC). It is gratifying that
the ADW result can be firmly based on the systematic chiral expansion of
low energy QCD Green functions. In chiral perturbation theory, one could in
principle calculate the corrections to equation~(\ref{e10}) by performing a two--loop
calculation while in Wolfenstein's method these could only be estimated. To
stress it again, the main ingredient to arrive at equation~(\ref{e10})
in CHPT is the
linear $t$--dependence in equation~(\ref{e6}). Since we are interested here
in a very small momentum transfer, like e.g.for the case of ordinary muon capture,
$t = -0.88M_\mu^2 \simeq -0.5 M_\pi^2$,
curvature terms of order $t^2$ have to be negligible. If one uses for example
the dipole parameterization for the axial form factor,
the $t^2$--term amounts to a 1.3$\%$ correction to the one
linear in $t$.
The masses $m$, $M_\mu$ and $M_\pi = M_{\pi^+}$ are accurately known
and so are $F_\pi$ and $g_A$ \cite{PDG}.
The situation concerning the strong pion--nucleon coupling constant is less
favorable. The methodologically best determination based on dispersion theory
gave $g_{\pi N}^2 / 4 \pi = 14.28 \pm 0.36$ \cite{LB}, more recent
determinations seem to favor smaller values \cite{AWP}.
We use here $g_{\pi N} = 13.10 \pm 0.35$. The value for the axial
radius has been discussed in section~\ref{sec:axdata}, we use here the
number obtained from neutrino scattering.
Putting pieces together, we arrive at 
\begin{equation}
g_P  = (8.74 \pm 0.23)  - (0.48 \pm 0.02) = 8.26 \pm 0.16~.
\label{e11}
\end{equation}
The uncertainties  stem from the range of $g_{\pi N}$
and from the one for $\langle r_A^2 \rangle$ 
for the first and second term, in order. For the
final result on $g_P$, we have added these uncertainties in quadrature. A
measurement with a 2$\%$ accuracy of $g_P$, as intended by the upcoming
experiment \cite{proposal},  could therefore cleanly separate
between the pion pole contribution and the improved CHPT result. This would
mean a significant progress in our understanding of this fundamental
low--energy parameter since the presently available determinations have too
large error bars to disentangle these values, see section~\ref{sec:gpdata}. In fact,
one might turn the argument around and eventually use a precise determination
of $g_P$ to get an additional determination of the strong pion--nucleon
coupling
constant which has been at the center of much controversy over the last years.
To summarize, we have shown that the chiral Ward identities allow
to predict the induced pseudoscalar coupling constant entirely in terms of
well--determined physical parameters within a few percent accuracy. As already
noted by Wolfenstein \cite{WOL}, an accurate empirical determination of this
quantity  therefore poses a stringent test on our understanding of the
underlying dynamics which is believed to be realized in the effective
low--energy field theory of QCD.

\subsection{Chiral expansion of ordinary muon capture}
\label{sec:OMC}
As detailed in section~\ref{sec:gpdata}, measurements of OMC
on liquid hydrogen allow one to extract a value for the induced
pseudoscalar coupling constant that agrees with theoretical expectations.
The theoretical description of the hadronic physics involved is fairly
transparent, we follow reference~\cite{BHM} and refer to that paper for 
many details. However, as stressed in particular by Ando, Myhrer and
Kubodera~\cite{AMK}, there are some difficulties with the accepted
picture of translating the atomic rates into the one for liquid
hydrogen. This topic will be taken up at the end of this section.
First, we consider the OMC (atomic) decay rates calculated to third order
in the chiral expansion and also to third order in the so--called
``small scale expansion'', which is an effective field theory
with explicit $\Delta (1232)$ degrees of freedom (for a detailed
introduction and review, see~\cite{HHK}).
In the Fermi approximation of a static $W_\mu^-$ field,
i.e. the gauge boson propagator is reduced to a point interaction
(since the typical momenta involved are much smaller than the $W$ mass),
\begin{eqnarray}
{\cal M}_{\mu^-p\rightarrow \nu_\mu n}&=&{\cal M}^{\rm OMC} =
                \langle \nu_\mu|W_\mu^+|\mu\rangle
                \,i\,\frac{g^{\mu\nu}}{M_W^2}\left[\langle n|V_\nu^-|p\rangle -
                \langle n|A_\nu^-|p\rangle\right]~,
\end{eqnarray}
where the vector current matrix element $\langle n|V_\nu^-|p\rangle$ (the so-called
vector correlator) is parameterized in terms of the  isovector vector Dirac and Pauli
form factors $F_{1,2}^v$, which are known empirically (see e.g. \cite{MMD,HMD})
and have also been studied in baryon CHPT~\cite{BKKM,BFHM,BKff}. The axial correlator
contains the desired dependence on the induced pseudoscalar form factor.
Introducing the Fermi constant $G_F$,
the square of the {\em spin-averaged} invariant matrix element is
defined via
\begin{equation}
\frac{1}{4}\sum_{\sigma\sigma^\prime s s^\prime}|{\cal M}^{\rm OMC}|^2
= \case{1}{2} \, G_F^2\, V_{ud}^2 \, L_{\mu\nu}^{(a)}H^{\mu\nu}_{(a)} \; .
\end{equation}
with $V_{ud} =0.974$ the pertinent CKM matrix element and
the explicit forms of the symmetric leptonic $L_{\mu\nu}$, see also 
equation~(\ref{Lnu}), and the hadronic tensor $H^{\mu\nu}$ are 
given in \cite{BHM}.  Approximating now
the muonic atom wave--function dependence by its value at the origin
(thus neglecting the small third order effects due to the form factors),
the $1s$ Bohr wave function of the muonic atom reads
\begin{equation}
\Phi(0)_{1s}= \frac{\alpha^{3/2}\mu^{3/2}}{\sqrt{\pi}}\; ,
\end{equation}
with the reduced mass $\mu={(m M_{\mu^-}})/{(m + M_{\mu^-})}$ and
$\alpha$ the fine--structure constant.
We can therefore calculate the spin--averaged rate of ordinary muon capture via
\begin{eqnarray}
\fl
\Gamma_{\rm OMC}&=&|\Phi(0)_{1s}|^2 \int\frac{d^3r^\prime}
               {(2\pi)^3 J_n}\,\frac{d^3 l^\prime}{(2\pi)^3 J_\nu}
               \left(2\pi\right)^4\delta^4\left(r+l-r^\prime-l^\prime\right)
               \frac{1}{4}
               \sum_{\sigma\sigma^\prime s s^\prime}|{\cal M}^{\rm OMC}|^2
\end{eqnarray}
with $J_\nu$, $J_n$ appropriate normalization factors.
Evaluating this to third order, $\Or(q^3)$, one obtains 
\begin{equation}\label{OMCrate}
\fl
\begin{tabular}{ccccc}
$\Gamma_{\rm OMC} = \biggl( \!\!\! $ & $\underbrace{247.0}$ & $\underbrace{-61.6}$
& $\underbrace{-3.8}$ & $ + \,\; \Or(1/m^3)\biggr)\times
                          {\rm s}^{-1} = 181.7\times {\rm s}^{-1}$~,
                           \\
& $\Or(q)$ &  $\Or(q^2)$&  $\Or(q^3)$ &
\end{tabular}
\end{equation}
which shows a very fast convergence. Retaining the delta only reduces
the small third order correction by 5\%.
The reason for the nice stability of perturbative calculations for 
OMC in the physical world of small finite quark masses is of course
the fact that contributions of order $n$ are suppressed by
$\left(M_i/\Lambda_\chi\right)^{n-1}$, with $i=\pi,\mu$ and
$\Lambda_\chi\sim m \sim 1$~GeV. This spin--averaged OMC scenario 
is mostly of theoretical interest. In nature the weak interactions
show a very strong spin--dependence, which
leads to quite different decay rates depending on whether the captured 1s muon
forms a singlet or a triplet spin-state with the proton \cite{Opat}, where the singlet
is the
usual state $(1/\sqrt{2})(|\uparrow,\downarrow \rangle
- |\downarrow, \uparrow\rangle)$ in terms of the muon and proton spins and the
triplet accordingly, see also figure~\ref{fig:omc} and the discussion in
section~\ref{sec:gpdata}. Although the hyperfine splitting between
the two levels ``only'' amounts to 0.04~eV, the occupation numbers of the levels
due to thermal and collision induced processes tend to be far from statistical
equilibrium. In order to make any contact with experiment, 
the singlet/triplet rates need to be calculated separately.
For the total capture rates of singlet and the triplet states in the muonic
atom, one find the following
decomposition into leading, next--to--leading  and
 next--to--next--to--leading order pieces
\begin{eqnarray}
\Gamma^{\rm sing}_{\rm OMC} &=& (957 - 245\,{\rm GeV}/m + (30.4\,{\rm
  GeV}/ m^2
-43.17) + \Or(1/m^3))\times {\rm
  s}^{-1} \nonumber\\
&=& 687.4\times {\rm  s}^{-1}~,\nonumber\\
\Gamma^{\rm trip}_{\rm OMC} &=& (10.3 + 4.72\,{\rm GeV}/ m - 
(1.22\,{\rm GeV}/m^2 +1.00)+{\cal O}(1/m^3))
\times {\rm s}^{-1} \nonumber\\
&=& 12.9\times {\rm s}^{-1}~,
\label{Vero}
\end{eqnarray}
displaying the dramatic spin-dependence due to the V-A structure of the
weak
interactions in the Standard Model. These numbers correspond to a value of
$g_{\pi NN} =13.05$ and include the pion pole 
corrections, cf. equation~(\ref{e10}).
Since $G_P$ contributes negatively to the singlet rate a larger
value of $g_{\pi NN}$ leads to a smaller value for the rate: 
$\Gamma^{\rm sing}_{\rm OMC}=681.9\times {\rm  s}^{-1}$ for $g_{\pi NN}=13.4$. Similarly,
neglecting the pion pole corrections leads to 
$\Gamma^{\rm sing}_{\rm OMC}=676.1\times {\rm  s}^{-1}$.
In Equation~(\ref{Vero}) we have split 
the third order term (third and fourth terms
in parenthesis) into the contribution from the $1/m^2$ corrections and the
terms stemming from the various radii which lead to the $Q^2$--dependence of the
 form factors. It is very interesting
to note that these two contributions more or less cancels themselves in 
the case of the singlet term. It is thus extremely important to 
perform a consistent chiral expansion. The existing theoretical
predictions for these rates are collected in table~\ref{tab:omc}.
\begin{table}[htb]
\caption{Calculated atomic singlet and triplet rates  in s$^{-1}$ from
BHM~\protect\cite{BHM}, AMK~\protect\cite{AMK}, Primakoff~\protect\cite{Prima}
and Opat~\protect\cite{Opat}. Note that the latter two have used
smaller values for $g_A$ and that AMK used $g_{\pi N} = 13.4$.
}\label{tab:omc}
\begin{indented}
\item[]\begin{tabular}{@{}lllllll}
\br
 & BHM & BHM & AMK & AMK & Primakoff & Opat\\
 & NLO & NNLO& NLO & NNLO&           &     \\
\mr
$\Gamma^{\rm sing}$ & 711  & 687.4 & 695  & 722  & $664\pm 20$   & 634
\\
$\Gamma^{\rm trip}$ & 14.0 & 12.9  & 12.2 & 11.9 & $11.9\pm 0.7$ & 13.3 \\
\br
\end{tabular}
\end{indented}
\end{table}
For comparison, in a
relativistic Born  model one obtains the simple nice formula
for $\Gamma^{\rm sing}_{\rm OMC}$ \cite{san}:
\begin{eqnarray}
\fl
\Gamma^{\rm sing}_{\rm OMC}&\sim&\left(6.236\,F_1^v(q_0^2)+0.5513\,F_2^v(q_0^2)
+16.44\, G_A(q_0^2) -0.2834\,G_P(q_0^2) \right)^2   \label{san}\nonumber \\
\fl
&\sim&683\times {\rm s}^{-1}
\end{eqnarray}
where $q_0^2= -0.88 M_\mu^2$ is the momentum transfer corresponding
to muon capture at rest (again, a rescaling of this formula to the
present day value of $g_A$ has been performed).
This formula though very appealing should not be used, being in contradiction with the
modern viewpoint of power counting. The good agreement with the CHPT result
is purely accidental.
The problem arises from the fact that 
$\Gamma^{\rm sing}_{\rm OMC}$  is a rather sensitive quantity as can be seen
for example in equation~(\ref{san}). Indeed the terms proportional to  
$F_2^v(q_0^2)$ and $G_P(q_0^2)$ are of the same order of magnitude
but have different signs so they have a tendency to cancel each other
rendering the values of $\Gamma^{\rm sing}$ rather sensitive to the
exact values of these two quantities. 
So far we have only considered OMC for the case of muonic atoms. For the
case of a liquid hydrogen target one also has to take into account the possibility of
muon capture in a muonic molecule $p\mu p$, which can be formed via the reaction
$p\mu+pep\rightarrow p\mu p+e+124\,$eV. In such a molecule
the muon can be found in the ortho $(O)$ (spin of the protons parallel) or para $(P)$
(spin of the protons antiparallel) spin state relative to its two accompanying
protons. The decay rates of these molecular states can be calculated from the singlet/triplet
rates of the muonic atom via
\begin{eqnarray}
\Gamma_P &=&2\gamma_P \frac{1}{4}(3\Gamma_{\rm trip} + \Gamma_{\rm sing} )~,
\label{para} \\
\Gamma_O &=&p_{1/2}\,\Gamma_{1/2}+p_{3/2}\,\Gamma_{3/2}~, \label{ortho}
\end{eqnarray}
with $\gamma_P\;(\gamma_O)$ denoting the ratio of the probability of finding the
negative muon at the point occupied by a proton in the para-muonic
(ortho-muonic)
molecule and the probability of finding the negative muon at the origin in the
muonic atom.
The wavefunction corrections are taken to be $\gamma_O=0.500,\;\gamma_P=0.5733$ \cite{WP}.
We note that the para molecular state is often referred to as the
statistical mixture, as it corresponds to the naively expected
occupation  numbers of the muonic
atom. For a precise calculation of the ortho molecular state on the other hand one
first has to know the exact probabilities $p_{1/2},p_{3/2}$ for the muonic molecule
being in a total spin S=(1-1/2)=1/2 or
a total spin S=(1+1/2)=3/2 state, with $p_{1/2}>0.5$ \cite{WeinO}, see
also figure~\ref{fig:omc}.
The corresponding decay rates are given by
\cite{Prima,WeinO}
\begin{eqnarray}
\Gamma_{1/2}&=&2\gamma_O\left(\frac{3}{4}\Gamma_{\rm sing}+\frac{1}{4}
\Gamma_{\rm trip}\right)\nonumber\\
\Gamma_{3/2}&=&2\gamma_O\;\Gamma_{\rm trip}
\end{eqnarray}
Theoretical calculations of the spin-effects in the muonic molecule
\cite{Bakalov} suggest $p_{1/2}\approx 1,\;p_{3/2}\approx 0$
which leads to the values:
\begin{eqnarray}
\Gamma_P^{\rm OMC} &=& 208\times {\rm s}^{-1}\nonumber\\
\Gamma_O^{\rm OMC} &=& 493 \cdots 519   \times {\rm s}^{-1}\; .\label{orthosimple}
\end{eqnarray}
where the range given in $\Gamma_O^{\rm OMC}$ corresponds to $0.95 \leq p_{1/2}
\leq 1$. We have allowed here for a 5\% uncertainty in the occupation
numbers to show the sensitivity of our results on this quantity.
This point was first made in reference~\cite{AMK}. 
Since $\Gamma_{\rm sing}\gg\Gamma_{\rm trip}$ for OMC, $\Gamma_O^{\rm OMC}$ turns out to be
roughly proportional to $ p_{1/2}$. We note that our number for
capture from the molecular ortho state agrees very well with the most recent
measurement $\Gamma_O^{\rm exp}=(531\pm33)\times {\rm s}^{-1}$ 
\cite{Bardin}. We stress again that a direct use of the liquid
transition formula {\`a} la reference~\cite{Bakalov} is only
meaningful if one accounts for the time structure of the beam in the
experiment, see also \cite{AKMt}.

\subsection{Chiral expansion of radiative muon capture}
\label{sec:RMC}
The pioneering TRIUMF RMC result, cf. equation~(\ref{TRIUMFv}), spurred a lot 
of theoretical activity. While radiative
muon capture had already been calculated in phenomenological tree
level models a long time ago, see e.g.~\cite{BF2,Opat,Beder,Fear,BF1},
heavy baryon chiral perturbation theory was also used at tree level including
dimension two operators~\cite{meiss} and to one loop order~\cite{AM}. The
resulting photon spectra are not very different from the ones obtained
in the phenomenological models, the most striking feature being the
smallness of the chiral loops and polynomial third order corrections~\cite{AM}, 
hinting towards a good convergence of
the chiral expansion. At present, the puzzling result from the TRIUMF
experiment has not fully been explained, but it appears now 
that the discrepancy does not come only from the strong interactions but
rather is also related to the distribution of the various spin states of the
muonic atoms, as detailed below~\cite{BHM,AMK}. 
It is therefore mandatory to sharpen the theoretical
predictions for the strong as well as the non--strong physics entering
the experimental analysis. Therefore,  RMC was also analyzed in the framework
of the small scale expansion~\cite{HHK}, which allows to
systematically include the $\Delta$ resonance into the effective field theory.
Although Ando and Min \cite{AM} have already shown that the RMC process
possesses a well behaved chiral expansion up to N$^2$LO, it has been noted quite
early \cite{Nimai} that one should reanalyze
RMC in a chiral effective field theory with explicit $\Delta$ degrees of
freedom.
This is due to the fact that the $\Delta$-resonance lies quite close to the nucleon
and therefore, in a delta-free theory like baryon CHPT, could lead
to unnaturally large higher order contact interactions which would spoil 
the seemingly good chiral convergence. Stating this in the language of (naive) dimensional
analysis, it suggests the possibility of corrections of the order
of 30\% due to the small nucleon--delta mass splitting, $M_\mu
/(m_\Delta - m) \sim 3 M_\mu / m$. Such a study including explicit delta degrees
of freedom was performed in~\cite{BHM}. It should also be mentioned that 
phenomenological models have claimed for a long time that the
$\Delta$ contribution does not exceed 8\% in the photon spectrum for photon
energies above 60~MeV~\cite{BF1}.  We will now proceed to discuss general
features of the chiral and the small scale expansion for RMC and then 
critically reexamine the way the TRIUMF experiment was analyzed.

\subsubsection{General results}
\label{sec:genr}
\begin{figure}[htb]
\centerline{
\epsfysize=6cm 
\epsffile{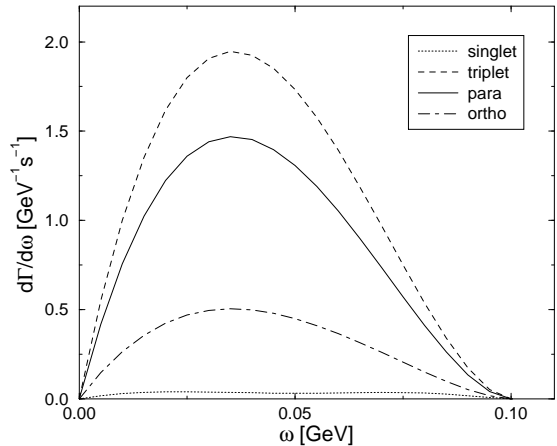}
}
\vskip 0.1cm 
\caption[]{
Photon spectra for RMC for the singlet, triplet, para (statistical) and
ortho states of the $\mu-p$ system calculated
to second order in the small scale expansion.
}\label{fig:spec}
\end{figure}

In the static approximation for the W--boson, the pertinent matrix element
for RMC decomposes into two terms,
\begin{eqnarray}\label{MRMC}
\fl
{\cal M}_{\mu^-p\rightarrow \nu_\mu n\gamma}&=&\langle
\nu_\mu|W_\mu^+|\mu\rangle
                \,i\,\frac{g^{\mu\nu}}{M_W^2}\left[
                \langle n|{\cal T} \,V\cdot\epsilon^\ast V_\nu^- |p\rangle
                -\langle n|{\cal T} \,V\cdot\epsilon^\ast A_\nu^- |p\rangle
                \right] \nonumber \\
\fl
             & &+\langle \nu_\mu\,\gamma|W_\mu^+|\mu\rangle
                \,i\,\frac{g^{\mu\nu}}{M_W^2}\left[\langle n|V_\nu^-|p\rangle -
                \langle n|A_\nu^-|p\rangle\right]~,
\end{eqnarray}
where the first one contains  the vector--vector (VV) and
vector--axial (VA) correlator. The hadronic part of the second term
is, of course,  identical to the OMC matrix--element.
Following ~\cite{BHM}, we give these to second
order in the chiral and the small scale expansion. The third order
terms have been worked out by Ando and Min \cite{AM} and found to be
small. Working in the  Coulomb gauge  for the photon
and making use of the transversality condition $\epsilon^\ast\cdot
k=0$, one finds (the pertinent momenta were already given in equation~(\ref{RMCdef}))
\begin{eqnarray}
\fl
\langle n|{\cal T} \,V\cdot\epsilon^\ast V_\mu^- |p\rangle^{(2)}& = -i\,
                             \frac{g_2 V_{ud}\,
                             e}{\sqrt{8}}\,\bar{n} (r^\prime)\left\{
                             \frac{1+\kappa_v}{m}\left[S_\mu,S
                             \cdot\epsilon^\ast
                             \right]-\frac{1}{2 m}\,\epsilon^\ast_\mu
                             \right. \nonumber \\
\fl
&  \left.
   +\frac{1}{m \omega}\,v_\mu \left[
   \left(1+\kappa_v\right)[S\cdot\epsilon^\ast,S\cdot k] -
   \epsilon^\ast\cdot r\right]+ \Or(1/m^2)
   \right\} p (r)~, \\
\fl
\langle n|{\cal T} \,V\cdot\epsilon^\ast A_\mu^- |p\rangle^{(2)}&= -i\,
                             \frac{g_2 V_{ud}\,
                             e}{\sqrt{8}}\,\bar{n}_ (r^\prime) \times \nonumber
\\
\fl
& \left\{2\,R\,g_A\,\frac{S\cdot(r^\prime-r)}{
   (r^\prime-r)^2-M_{\pi}^2}\times\left[\frac{2\,\epsilon^\ast\cdot(l-l^
   \prime)\,
   (l-l^\prime)_\mu}
   {(l-l^\prime)^2-M_{\pi}^2}-\epsilon^\ast_\mu\right]\right. \nonumber \\
\fl
& - R\,\frac{g_A}{m}\,\frac{\left(v\cdot r^\prime-v\cdot r\right)
   S\cdot(r+r^\prime)}{(r^\prime-r)^2-M_{\pi}^2}
   \times\left[\frac{2\,\epsilon^\ast\cdot(l-l^\prime)\,(l-l^\prime)_\mu}
   {(l-l^\prime)^2-M_{\pi}^2}-\epsilon^\ast_\mu\right]
   \nonumber \\
\fl
& -2\,R\,g_A\left[1+\frac{v\cdot l-v\cdot l^\prime}{2 m}\right]\frac{S\cdot
   \epsilon^\ast\,(l-l^\prime)_\mu}{(l-l^\prime)^2-M_{\pi}^2} +
\frac{g_A}{m}\,S\cdot
   \epsilon^\ast\,v_\mu \nonumber \\
\fl
& +\frac{g_A}{m}\left[\frac{(2+\kappa_s+\kappa_v)\,
   S^\alpha\,[S\cdot\epsilon^\ast,S\cdot
k]}{\omega}+\frac{(\kappa_v-\kappa_s)\,
   [S\cdot\epsilon^\ast,S\cdot k]\,S^\alpha}{\omega}\right.\nonumber \\
\fl
&  \left.
   -\frac{2\,S^\alpha\epsilon^\ast\cdot r}{\omega}\right]
   \times\left[g_{\mu\alpha}-R\,\frac{(l-l^\prime)_\alpha(l-
   l^\prime)_\mu}{(l-l^\prime)^2-M_\pi^2}\right] \nonumber \\
\fl
& +\frac{g_{\pi N\Delta}b_1}{3\, m}\left[\frac{2\Delta\,[k^\alpha
S\cdot\epsilon^\ast
   -\omega\, v^\alpha S\cdot\epsilon^\ast-\epsilon^{\ast\,\alpha}S\cdot
k]}{\Delta^2-
   \omega^2}-\frac{4\,S^\alpha[S\cdot\epsilon^\ast,S\cdot
k]}{3\,(\Delta+\omega)}
   \right. \nonumber \\
\fl
&  \left. \left.
    +\frac{4\,[S\cdot\epsilon^\ast,S\cdot
k]\,S^\alpha}{3\,(\Delta-\omega)}\right]
   \times\left[g_{\mu\alpha}-\frac{(l-l^\prime)_\alpha(l-l^\prime)_\mu}{(l-l^
   \prime)^2-
   M_\pi^2}\right] + \Or(1/m^2)\right\} p (r)~,
\nonumber \\ \fl &
\label{VAcorr}
\end{eqnarray}
with $\omega = v \cdot k$, $S^\mu$ the covariant nucleon spin--vector,
$v_\mu$ the nucleons four--velocity (see e.g. reference~\cite{BKKM}
for a more detailed discussion of the underlying heavy baryon formalism) 
and $R=1$ in QCD. Furthermore,
$\kappa_s$ is the nucleon isoscalar anomalous magnetic moment, $\Delta
= m_\Delta -m$ the delta--nucleon mass splitting, $g_{\pi N \Delta}$
and $b_1$ are the leading $\pi N \Delta$ and $\gamma N \Delta$
coupling  constants. Numerically, $g_{\pi N \Delta} \times b_1 = 1.05
\times 12 = 12.6$~\cite{HHK}. The SU(2) gauge coupling $g_2$ is
related to the Fermi constant via $G_F = g_2^2 \sqrt{2}/(8M_W^2)$.
Note that the vector--vector correlator is free of delta
effects at next--to--leading order, i.e. the leading $\Delta$(1232)
effect only appears in the vector--axial correlator. The latter terms
constitute the difference between the chiral and the small scale
expansions for RMC evaluated to NLO. The factor $R$
multiplying the Born term contributions proportional to the induced
pseudoscalar form factor has been introduced for the later
discussion. Note again that these expressions do not contain the
full form factor $G_P (t)$ but only its leading part (the CHPT
correction given in equation~(\ref{e9}) does not appear at this order).
One can easily check from the 
continuity equations satisfied by the correlators 
that gauge invariance is satisfied in the above equations.
We are now in the position to calculate the decay rates.
Consider first the spin--averaged case. The square of the
matrix--element equation~(\ref{MRMC})  can be written as a sum of
four terms, with both photons coming either from the hadronic or the
leptonic side and two mixed terms, i.e.
\begin{equation}
\fl
\frac{1}{4}
\sum_{\sigma\sigma^\prime s s^\prime\lambda\lambda^\prime}|{\cal
  M}^{\rm RMC}|^2 =
\frac{e^2 G_F^2 V_{ud}^2}{2}\left[L_{\mu\nu}^{(a)}H^{\mu\nu}_{(d)}
+\left(\sum_{\lambda\lambda^\prime}L_{\mu\nu}^{(b)}H^{\mu\nu}_{(c)}+
L_{\mu\nu}^{(c)}H^{\mu\nu}_{(b)}\right)
+L_{\mu\nu}^{(d)}H^{\mu\nu}_{(a)}\right]~,
\end{equation}
with $\lambda , \lambda'$ the photon helicities.
Explicit expressions for the various tensors are not given here
because they are lengthy and not illuminating. 
The total decay rate is given by:
\begin{eqnarray}\label{RMCrate}
\fl
\Gamma_{\rm tot}
= \frac{ |\Phi(0)_{1s}|^2 }{16\pi^4} \,
 \int_0^\pi  \sin\theta d\theta \, \int_0^{\omega_{\rm max}}
 d\omega \,  \omega \, l_0'\,
\biggl( 1 - & \biggl( \frac{m_\mu - \omega (1 -
  \cos\theta)}{m}\biggr)\biggr)\, \nonumber \\
& \times \frac{1}{4}\, \sum_{\sigma\sigma^\prime s s^\prime\lambda\lambda^\prime}
\,\, |{\cal M}^{\rm RMC}|^2~, 
\end{eqnarray}
with $\omega = k_0$ the photon energy (note that in this and the following subsection
we exceptionally use the symbol $\omega$ for the photon energy to facilitate the
comparison with the literature). 
The direction of the photon defines
the z--direction and $\theta$ in equation~(\ref{RMCrate}) is the polar angle of
the outgoing lepton with respect to this direction. The maximal photon energy
is given by
\begin{equation}
\omega_{\rm max} = M_\mu\,\biggl( 1 + \frac{M_\mu}{2m}\biggr) \,
\biggl( 1 + \frac{M_\mu}{m}\biggr)^{-1} ~.
\end{equation}
Furthermore, the energy of the outgoing lepton follows from energy
conservation,
\begin{equation}\label{l0}
l_0'  = M_\mu  - \omega -\frac{M_\mu^2}{m} + \frac{\omega(1-\cos\theta)(M_\mu
- \omega)}{m} + \Or(1/m^2)~.
\end{equation}
First we discuss the (academic) spin-averaged RMC scenario, which allows for
a comparison with previous calculations:
\begin{eqnarray}
\fl
\Gamma^{\rm RMC}_{\rm spin av.}&=&\left(66.0+18.7+{\cal O}(1/m^2)\right)
\times10^{-3}\,{\rm s}^{-1}=84.7\times10^{-3}\,{\rm s}^{-1}\;{\rm (CHPT)}\nonumber\\
\fl
\Gamma^{\rm RMC}_{\rm spin av.}&=&\left(66.0+20.4+{\cal O}(1/m^2)\right)
\times10^{-3}\,{\rm s}^{-1}=86.4\times10^{-3}\,{\rm s}^{-1}\;{\rm (SSE)} .
\end{eqnarray}
Both the CHPT and the SSE results suggest a good convergence for the chiral
expansion, as expected from dimensional analysis. Note that the leading order
capture rates in both calculations are identical,
as $\Delta$(1232) related effects only start at sub-leading order. Our leading
order result also agrees with the  calculation of reference~\cite{meiss}. 
For the case of muonic atoms we obtain the following decay rates in the
singlet/triplet channel (note the reversal of
the relative size of the singlet to triplet contribution as compared
to the OMC case),
\begin{eqnarray}
\fl
\Gamma^{\rm RMC}_{\rm sing} &= \left(12.7-18.7\,{\rm GeV}/m+{\cal O}(1/m^2)\right)
\times10^{-3}\,{\rm s}^{-1}=3.10\times10^{-3}\,{\rm s}^{-1}\;{\rm (CHPT)}\nonumber\\
\fl
\Gamma^{\rm RMC}_{\rm sing}&= \left(12.7-18.3\,{\rm GeV}/m+{\cal O}(1/m^2)\right)
\times10^{-3}\,{\rm s}^{-1}=2.90\times10^{-3}\,{\rm s}^{-1}\;{\rm (SSE)}
\nonumber \\
\fl
\Gamma^{\rm RMC}_{\rm trip}&= \left(119-3.86\,{\rm GeV}/m+{\cal O}(1/m^2)\right)
\times10^{-3}\,{\rm s}^{-1}=112\times10^{-3}\,{\rm s}^{-1}\;{\rm (CHPT)}\nonumber\\
\fl
\Gamma^{\rm RMC}_{\rm trip}&= \left(119-1.80\,{\rm GeV}/m+{\cal O}(1/m^2)\right)
\times10^{-3}\,{\rm s}^{-1}=114\times10^{-3}\,{\rm s}^{-1}\;{\rm (SSE)}
\end{eqnarray}
It should be stressed that for the total numbers given the
kinematical factors were not expanded in powers of $1/m$ since in case
of the small singlet, the contribution from the terms starting at
order  $1/m^2$ can not be neglected.
Next, we address the complications for RMC due to the presence of muonic
molecules in the liquid hydrogen target. According to equation~(\ref{para}), we can
easily determine the capture rate from the molecular para state
\begin{equation}
\Gamma^{\rm RMC}_{P} = 85.2 \,[86.4] \,\times10^{-3}\,{\rm s}^{-1}\;{\rm
  (CHPT \, [SSE])}~.
\end{equation}
Consider now the molecular ortho state, which turns out to dominate in the recent RMC
experiment from TRIUMF \cite{TRIUMF1,TRIUMF2}. One obtains for $p_{1/2}=1$:
\begin{equation}
\Gamma^{\rm RMC}_{O} =
30.4 \, [30.8] \,\times10^{-3}\,{\rm s}^{-1} \;{\rm (CHPT \, [SSE])}~,
\end{equation}
i.e in both cases the delta effects are fairly small.
Due to the triplet dominance in RMC (as opposed to the singlet dominance 
in OMC) $\Gamma^{\rm RMC}_{O}$ is now roughly proportional to $(1-3/4p_{1/2})$ 
which leads to a big sensitivity of the RMC capture rate to the exact
occupation numbers of the relative molecular
sub-states. For example, a 5\% uncertainty in the
occupation numbers $p_{1/2}=0.95,\;p_{3/2}=0.05$ would lead to a 13\% change in
the ortho capture rate $\Gamma^{\rm RMC}_{O}\sim 35\times10^{-3}\,{\rm s}^{-1}$.
We will discuss the implications of this uncertainty when we compare our
results with the measured photon spectrum from TRIUMF in the next
subsection.  For the total capture rate in the TRIUMF experiment
\begin{equation}
\Gamma_{\rm RMC}^{H_2}=f_S\;\Gamma_{\rm sing}^{\rm RMC}+f_O\;\Gamma_O^{\rm RMC}+
f_P\;\Gamma_P^{\rm RMC}
\end{equation}
with $f_S=0.061,\,f_O=0.854,\,f_P=0.085$ \cite{TRIUMF1,TRIUMF2} one would obtain
$\Gamma_{\rm RMC}^{TRIUMF}=(34.3\;[34.8]+ \Or(1/m^2))\times 10^{-3}\,{\rm
s}^{-1}$ in HBCHPT [SSE]. This
leads to a relative branching ratio $Q_\gamma=\Gamma_{\rm RMC} / \Gamma_{\rm OMC}$
\begin{equation}
Q_\gamma =
 6.8 \, [6.9] \, \times 10^{-5} + \Or(1/m^2) ~~ {\rm (CHPT \,
 [SSE]\,)}~.
\end{equation}
Unfortunately the full relative branching ratio is not accessible in
experiment, as one has to use a severe cut on the photon energies due to strong
backgrounds. In the TRIUMF experiment only
photons with an energy $\omega>60\,$MeV were detected. We therefore now move on
to a discussion of the photon spectrum $d\Gamma / d\omega$.
It  can be obtained straightforwardly
from equation~(\ref{RMCrate}). 
The calculated photon spectra to second order in CHPT and SSE 
come out  very similar.  In figure~\ref{fig:spec}
we show the second order SSE  results for the singlet, triplet,
para and ortho states. 
The relative difference between the second order CHPT and  SSE
calculation for all states are of the order of a few percent, 
showing explicitly the small role of 
$\Delta$(1232) in RMC. These results 
are very similar to the ones found by Beder and Fearing~\cite{BF2,BF1}
for the spectra and the relative contribution from the spin--3/2
resonance, although their calculation is based on a very different
approach. Even the result for the small  singlet is comparable though
not identical to the one of Beder and Fearing. We conclude that the
strong interaction theory for RMC is under control.

\subsubsection{Discussion of the TRIUMF result}
\label{sec:TRIUMF}
\begin{figure}[htb]
\centerline{
\epsfysize=2.in 
\epsffile{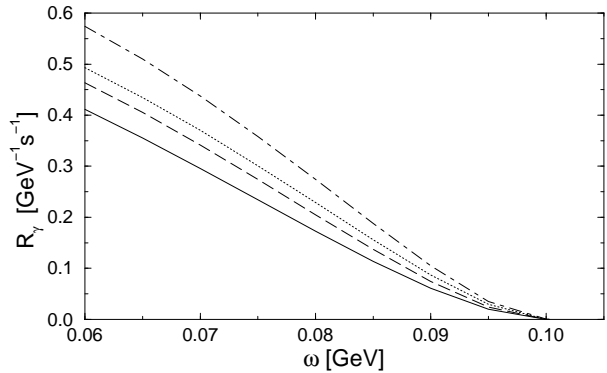}
}
\caption[]{
Photon spectra for RMC for the branching ratios of the singlet, ortho
and para states as used in the TRIUMF analysis. Solid line: Prediction
of the small scale expansion to order $\epsilon^2$. Dash--dotted line:
Same as the solid line but with $g_P$ scaled by a factor
$1.5$. Dashed line: Various small modifications as explained in the
text. Dotted line: Same as the dashed line but using the neutral
instead of the charged pion mass. 
}\label{fig:specR}
\end{figure}

The photon spectra just discussed  allow in
principle to determine the induced pseudoscalar form factor.
The TRIUMF result for $g_P$ is obtained by multiplying the terms
proportional to the pseudoscalar form factor with a constant denoted
$R$ (the momentum dependence is assumed to be entirely given by the pion
pole). The value of $R$ is then extracted using the model of Fearing \etal
to match the partial rate for photon energies larger than 60~MeV.
If we perform such a procedure, we get a similar shift in the partial
photon spectra. It is, however, obvious from the
analysis presented above that such a procedure is not legitimate. By artificially
enhancing the contribution $\sim g_P$ 
(to simulate this procedure,
we have introduced the factor $R$ in equation~(\ref{VAcorr}) and
similarly one must multiply $\langle n |A_\mu^-|p\rangle$ by $R$),
one mocks up a whole class of new contact and other terms not present in the
Born term
model. To demonstrate these points in a more quantitative
fashion, we show in figure~\ref{fig:specR} the partial branching fraction
in comparison to the one with $g_P$ enhanced by
a factor 1.5 and a third curve, which is obtained by increasing $g_P$
only by 15\% and slightly modifying some parameters related to the delta
contribution. This is shown by the dashed line and it shows
that such a combination of small effects can explain most (but not
all) of the shift in the spectrum. 
\noindent
This is further sharpened by using
now the neutral pion mass of 134.97~MeV instead of the charged pion
mass, leading to the dotted curve in figure~\ref{fig:specR}. Since the
pion mass difference is almost entirely of electromagnetic origin, one
might speculate that isospin--breaking effects should not be neglected.
 Furthermore, as
discussed above a slight change in the occupation numbers $p_{1/2}$
and $p_{3/2}$ would also lead to an increase in the ortho capture rate
which could close the gap between the empirical and theoretical
results. For example the dashed curve in figure~\ref{fig:specR}
would be moved from 0.42 to 0.48
while the dotted one would go from 0.50 to 0.54 with $p_{1/2}=0.95$
and  $p_{3/2}=0.05$ at $\omega = 60\,$MeV.
The situation is reminiscent
of the sigma term analysis, where many small effects combine to give
the sizeable difference between the sigma term at zero momentum
transfer and at the Cheng-Dashen point. 
Another point against the simple rescaling comes 
from OMC. Indeed if this rescaling holds for RMC it should also hold
for OMC. We thus have performed a similar calculation in OMC. Taking the
same value for R, one would obtain 
\begin{equation}
\fl
\Gamma_{\rm OMC}^{R=1.5} = 172.8 \times {\rm s}^{-1}, \quad  
\Gamma_{\rm OMC}^{\rm sing,\;R=1.5} = 634.6 \times {\rm s}^{-1}, \quad 
\Gamma_{\rm OMC}^{\rm trip,\;R=1.5} = 18.9 \times {\rm s}^{-1}, 
 \end{equation}
leading to $\Gamma^{\rm OMC,\;R=1.5}_O = 477 \times {\rm s}^{-1}$, which is lower than
the error bars of the experimental result from
reference~\cite{Bardin}. 
As expected, the   singlet and triplet capture rates are much more sensitive 
to the details of the interaction than the total rate. As a conclusion we note 
that the effect of enhancing the capture rates in RMC via setting $R=1.5$ 
leads to a strong reduction of the corresponding OMC rates leading to 
conflicts with the experimentally determined ortho capture rate.
To summarize this discussion, we have pointed out that two effects in
particular have to be investigated in more detail: (1) the occupation 
numbers of the atomic structure in muonic atoms/molecules
need to be carefully re-examined, and (2)
the N$^2$LO calculation should be redone including all isospin breaking effects
because of the sensitivity to the exact pion mass in the pion-pole 
contributions, for example. The sum of these small effects should explain the observed 
photon spectrum, as we believe that the proper hadronic/weak physics part
is well under control by now. A simple rescaling of the pseudoscalar coupling constant
should no longer be considered.

\section{Octet Goldberger-Treiman discrepancies}
\label{sec:GTR}
In section~\ref{sec:gp} we already encountered the Goldberger--Treiman discrepancy
(GTD), i.e. the deviation from the Goldberger--Treiman relation (GTR), which is exact 
in the chiral limit. This deviation is parameterized in terms of the quantity
$\Delta_\pi$ given by 
\begin{equation}
\Delta_{\pi} = 1 - \frac{m \, g_A}{F_\pi g_{\pi NN} (M_\pi^2)} 
= 1 - \frac{g_{\pi NN} (0)}{g_{\pi NN} (M_\pi^2)}~,
\end{equation}
using the GTR $m\, g_A(0) = F_\pi \, g_{\pi NN} (0)$. Inserting the PDG values for
$g_A$, $F_\pi$ and $m = (m_p + m_n)/2$, one obtains
\begin{equation}\label{GTDpi}
\Delta_{\pi} = 0.014 \, (0.040) ~~{\rm for}~~g_{\pi NN} = 13.05 \, (13.40)~.
\end{equation}
From naive dimensional analysis one can can estimate $\Delta_\pi \simeq
(M_\pi / \Lambda_\chi)^2 =  (M_\pi /
4\pi F_\pi)^2 \simeq 0.015$, so this favors the smaller pion--nucleon coupling
constant. Note, however, that if one chooses the $\rho$--meson mass as the relevant
hadronic scale, $\Lambda_\chi = M_\rho$,
this estimate increases to $(M_\pi / M_\rho)^2 \simeq 0.033$. 
One can further sharpen this argument by extending the analysis to
flavor SU(3) and also investigate consistency of the strange to light quark mass
ratio  as obtained from the Goldstone boson spectrum. This is based on the
observation that while in SU(2) the GTD is given in terms of one LEC, in chiral
SU(3) this quantity is given in terms of a sum of quark masses times an SU(3)
octet operator. This leads to a relation between various GTDs in the octet,
the so--called Dashen--Weinstein relation~\cite{DW}. This can be explored more
systematically in the framework of CHPT. We follow here
essentially reference~\cite{FSS} and the recent update given in \cite{goity} (see
also \cite{Barry}). In the baryon octet, one calculates matrix--elements of the
axial--currents $\bar{u} \gamma_\mu \gamma_5 d $ and $\bar{u} \gamma_\mu \gamma_5 s$
at zero momentum transfer,
\begin{equation}
\langle B' (p) | A_\mu | B (p) \rangle = \bar{u}_{B'} \, \left[
g_A^{B'B} \, \gamma_\mu \gamma_5 + \ldots \right] \, u_B~,
\end{equation}
and the pertinent axial coupling constants $g_A^{B'B}$ can be measured in semi--leptonic
hyperon decays ($\beta$--decays), $B \to B' \ell \nu_\ell$. Extending the analysis
of section~\ref{sec:gp} to the SU(3) case, one can derive a variety of GTR's and
the chiral corrections to these. We focus here on the three cases for which
sufficient empirical information for the corresponding strong meson-baryon
coupling constants  is available, i.e. $np$, $\Lambda p$ and
$\Sigma^- n$ transitions. To linear order in the quark masses, the chiral Ward
identity relating the divergence of the axial current to the pseudoscalar
density sandwiched between baryon states leads to
\begin{eqnarray}
(m_n+m_p) g_A^{np} &=  2F_\pi g_{\pi NN} & + (m_u+m_d) H^{np} (0) + \Or(m_q^2)~,
\nonumber \\
(m_\Lambda+m_p) g_A^{\Lambda p} &= 
-\sqrt{2}F_K g_{\Lambda} & + (m_u+m_s) H^{\Lambda p} (0) + \Or(m_q^2)~,
\nonumber \\
(m_{\Sigma^-} +m_n) g_A^{\Sigma^- n} &= 
{2}F_K g_{\Sigma} & + (m_u+m_s) H^{{\Sigma^-} n} (0) + \Or(m_q^2)~,
\end{eqnarray}
with $g_\Lambda = -g_{\Lambda p K^-}$ and $g_\Sigma = g_{\Sigma^- n K^-}/\sqrt{2}$.
Furthermore, the functions $H^{BB'} (t)$ are SU(3) generalizations of $h(t)$ in
equation~(\ref{e4}), given by
\begin{eqnarray}
\langle p | \bar{u} i \gamma_5 d |n \rangle & = \left[ H^{np} (t) + \pi-{\rm pole}
\, \right] \, \bar{u}_p i \gamma_5 u_n~, \nonumber \\
\langle p | \bar{u} i \gamma_5 s |\Lambda \rangle & = \left[ H^{\Lambda p} (t) 
+ K-{\rm pole} \, \right] \, \bar{u}_p i \gamma_5 u_\Lambda~, \nonumber \\
\langle n | \bar{u} i \gamma_5 s |\Sigma^- \rangle & = \left[ H^{{\Sigma^-} n} (t) 
+ K-{\rm pole} \, \right] \, \bar{u}_n i \gamma_5 u_{\Sigma^-}~.
\end{eqnarray}
Note that we are using the convention for the axial charges so that $g_A^{np}$ is 
positive. The $H$ functions admit a chiral expansion,
\begin{eqnarray}
H^{np} (0) = F + D + \Or(m_q) ~, \nonumber \\
H^{\Lambda p}(0) = -\sqrt{\case{2}{3}} \left(F + \case{1}{3}D \right)
+ \Or(m_q) ~, \nonumber \\
H^{{\Sigma^-} n}  (0) = -F + D + \Or(m_q) ~.
\end{eqnarray}
Here, $F=0.46$ and $D=0.80$ are SU(3) axial--vector couplings. 
Therefore we have three relations with two unknowns, from that we can deduce a
sum rule in terms of the GTD's ($\Delta_\pi, \Delta_K^\Lambda,  \Delta_K^\Sigma$)
and the quark masses, sometimes called the Dashen--Weinstein relation (DWR),
\begin{equation}\label{DWR}
\Delta_\pi = {F_K \over F_\pi} \, { \hat m \over
\hat m + m_s} \, \left(\sqrt{3} {g_{K\Lambda N} \over g_{\pi N N}} \Delta_K^\Lambda
+  {g_{K\Sigma N} \over g_{\pi N N}} \Delta_K^\Sigma
\, \right)~,
\end{equation}
in terms of the kaon discrepancies $\Delta_K^{\Lambda , \Sigma}$,
\begin{eqnarray} 
\Delta_K^\Lambda = 1 - {\sqrt{3} (m_p + m_\Lambda) \, g_A^{\Lambda p} (0) \over
2 \, F_K \, g_{K\Lambda N} (M_K^2)} ~, \nonumber\\
\Delta_K^\Sigma = 1 - { (m_n + m_{\Sigma^-}) \, (-g_A^{{\Sigma^-} n} (0)) \over
2 \, F_K \, g_{K\Sigma N} (M_K^2)} ~.
\end{eqnarray}
and $\hat m = (m_u + m_d)/2$ the average light quark mass.
There are two ways to explore the DWR. First, consider the quark mass ratio appearing
in equation~(\ref{DWR}) known, $2\hat m/(\hat m + m_s) = M_\pi^2 / M_K^2 \simeq 0.08$.
Note that higher order strong as well as electromagnetic corrections do not
change this leading order estimate appreciably, see e.g. \cite{leutm}.
Then one can ask the question whether three GTDs are mutually consistent, that is
whether the DWR is fulfilled. Another way of exploring this relation is to determine
the three discrepancies from data and then use the DWR to deduce the quark mass
ratio $\hat m /(\hat m +m_s)$. As input, we use the PDG values for the masses and 
axial coupling constants together with $g_{K\Lambda N} (M_K^2) = 13.7 \pm 0.4$
and $g_{K\Sigma N} (M_K^2) = 3.9 \pm 0.4$ as obtained from the analysis of
hyperon--antihyperon production at LEAR~\cite{NijmYY}. These numbers are close
to the expectations from flavor SU(3), $g_{K\Lambda N}  = 13.5$ and 
$g_{K\Sigma N} = 4.3$, respectively, and in rough agreement with the
analysis of  $\bar K$N scattering data given in reference~\cite{Martin}.
This leads to the following values of the kaon discrepancies
\begin{equation}\label{GTDK}
\Delta_K^\Lambda = 0.16 \pm 0.06~, \quad \Delta_K^\Sigma = 0.18 \pm 0.11~.
\end{equation}
These numbers conform to the expectations from SU(3) symmetry breaking,
$\Delta_K \simeq (M_K /\Lambda_\chi )^2 = 0.18$, but the theoretical uncertainties
might be larger as the ones given, see the discussion at the end of this section. 
Using now the quark mass ratio as deduced from the Goldstone boson masses and
the central values of the GTDs, one obtains $\Delta_\pi = 0.016$, which favors
the low value of the pion--nucleon coupling constant. 
If, on the other hand, one uses the empirical values of the
GTDs, see equations~(\ref{GTDpi},\ref{GTDK}), one obtains for the quark mass ratio
\begin{equation}
\frac{m_s}{\hat{m}}
= \left\{ \begin{array}{rl}
22.4 \pm 6.8 & {\rm for}~~g_{\pi NN} = 13.05\\
7.4 \pm 3.3 & {\rm for}~~g_{\pi NN} = 13.40
\end{array}\right. ~,
\end{equation}
i.e. the smaller value of $g_{\pi NN}$ is consistent with the standard scenario
of chiral symmetry breaking,  that is a large value of the quark condensate,
$B = \langle 0|\bar{q}q|0\rangle / F_\pi^2 \simeq 1.5\,$GeV. One could be content
with this consistency since the so determined quark mass ratio is in good agreement
with the one obtained from analyzing the Goldstone boson mass spectrum based on the
large condensate scenario. However, as already stressed in \cite{FSS},
a better and more reliable determination of the strong coupling constants
$g_{K\Lambda N}$ and $g_{K\Sigma N}$ is called for. In this context
it is worth to stress that extraction of these couplings from kaon electroproduction data
usually leads to very different values, see e.g. \cite{AS} and 
the discussion in reference~\cite{HMX}.
Note, however, that most of these kaon electroproduction  analyses are based on 
simple Born terms (including
resonance excitations and form factors) and are therefore plagued with large
theoretical uncertainties. In the light of the new and coming data on kaon 
electropoduction off protons and deuterium from ELSA and CEBAF,
it would  be very valuable to apply the theoretical
framework of references~\cite{OM1,OM2}, which combines chiral perturbation theory,
unitarity and coupled channel dynamics (see also \cite{MuX,ValX}). The presently existing
uncertainties for these coupling constants inhibit a precise test 
of the chiral symmetry breaking pattern of QCD.

\section{Summary and outlook}
\label{sec:summ}

The nucleon as probed with the weak axial current can be parameterized in
terms of two form factors, the axial, $G_A (t)$, and the induced
pseudoscalar, $G_P(t)$, one.
In this short review, we have shown that precise theoretical methods 
based on the symmetries of QCD exist for extracting these fundamental
observables from experiment. The axial form factor can be well
described by a dipole, $G_A (t) = (1 - t/M_A^2)^{-2}$.
The dipole mass $M_A$ can be translated into an
axial root--mean--square radius of $\langle r_A^2 \rangle^{1/2} = 0.67 \pm 0.01\,$fm.
This value is consistently obtained from (anti)neutrino scattering off protons (or
light nuclei) and charged pion electroproduction off protons. Clearly, more 
precise electroproduction data in the threshold region would be
welcome to further pin down this quantity. The induced pseudoscalar
form factor is dominated by the pion pole, but the small corrections
to this leading order result have been calculated. Existing data from
ordinary muon capture are consistent with these theoretical
expectations but have too large error bars to cleanly test
the chiral dynamics of QCD. We have argued that the result of the pioneering
TRIUMF  radiative muon capture experiment should be taken {\it cum
  grano salis} due to some assumptions in the analysis that are
inconsistent with the power counting underlying the effective field
theory of the Standard Model. However, it is fair to say that more
theoretical as well as experimental effort is needed for drawing a
final conclusion. The momentum--dependence of the
induced pseudoscalar form factor is
dominated by the pion pole which has only been tested in one
electroproduction experiment so far.
Also, a better determination of strong
kaon-hyperon-nucleon coupling constants $g_{K\Lambda N}$ and
$g_{K\Sigma N}$ would allow for an additional stringent
bound on the light to strange quark mass ratio, based on the
derivations from the octet Goldberger--Treiman relations. All this shows that
 precision experiments in hadron and nuclear physics indeed help to
unravel the mysteries of QCD at energies where one really has to deal
with {\it strong} interactions. Therefore, more pion and kaon
electroproduction as well as muon capture experiments are called for. 

\ack
We are grateful to Norbert Kaiser and Thomas Hemmert for fruitful collaborations,
Fred Myhrer for sharing his insight into the theory of muon capture and
Simon \v Sirca for detailed explanation of the MAMI data and analysis. Useful
communications from Barry Holstein, Jacques Martino and Claude
Petitjean  are acknowledged. We thank Josef Speth for initiating this project.
We also thank  the Deutsche Forschungsgemeinschaft and
the ELSA/MAMI Schwerpunktprogramm for funding the workshop on ``Hadron
Form Factors'' in Bad Honnef in April 2001, which triggered this write-up.
This work was supported in part by the NSF, grant 34201.

\end{document}